\documentclass[a4paper, 10pt, leqno]{amsart}

\usepackage{xcolor}
\definecolor{lgray}{RGB}{240,240,240}
\definecolor{webgreen}{rgb}{0,.5,0}
\definecolor{webbrown}{rgb}{.6,0,0}
\definecolor{RoyalBlue}{cmyk}{1, 0.50, 0, 0}
\usepackage[colorlinks=true, breaklinks=true, urlcolor=webbrown, linkcolor=RoyalBlue, citecolor=webgreen, backref=page]{hyperref}

\usepackage{mathtools}
\mathtoolsset{showonlyrefs}
\usepackage{epsfig, graphicx} 
\usepackage{verbatim, setspace}
\usepackage{newtxtext,newtxmath}

\usepackage{mathabx}
\usepackage{pifont}
\usepackage{subcaption}
\usepackage{array}
\usepackage{multirow}
\usepackage{titlesec}
\usepackage{titletoc}

\titlecontents{section} 
[0pt]                 
{\bfseries}            
{\thecontentslabel\quad} 
{}                    
{\titlerule*[0.5em]{.}\contentspage} 

\titlecontents{subsection} 
[5pt]                 
{}            
{\thecontentslabel\quad} 
{}                    
{\titlerule*[0.5em]{.}\contentspage} 

\titlecontents{subsubsection} 
[10pt]                 
{}            
{\thecontentslabel\quad} 
{}                    
{\titlerule*[0.5em]{.}\contentspage} 

\titlecontents{paragraph} 
[15pt]                 
{}            
{\thecontentslabel\quad} 
{}                    
{\titlerule*[0.5em]{.}\contentspage} 

\titlespacing*{\section}{0pt}{5.5ex plus 1ex minus .2ex}{4.3ex plus .2ex}

\numberwithin{equation}{section}

\titleformat{\section}{\centering\normalfont\large\scshape}{\thesection}{1em}{}

\titleformat{\subsection}{\normalfont\normalsize\bfseries}{\thesubsection}{1em}{}

\titleformat{\subsubsection}{\normalfont\normalsize\bfseries}{\thesubsubsection}{1em}{}

\titleformat{\paragraph}{\normalfont\normalsize\it}{\theparagraph}{1em}{}

\newtheorem{theorem}{Theorem}
\newtheorem*{DZthm}{Theorem (Deift-Zhou)}

\newtheorem{lemma}[theorem]{Lemma}

\newtheorem{RHP}{Riemann-Hilbert Problem}

\newcommand{\D}		{\mathbb{D}}
\newcommand{\R}		{\mathbb{R}}
\newcommand{\C}		{\mathbb{C}}

\newcommand{\Z}		{\mathbb{Z}}

\newcommand{\re}{\mathrm{Re}}
\newcommand{\im}{\mathrm{Im}}

\renewcommand{\arg}{\mathrm{arg}}
\renewcommand{\det}{\mathrm{det}}
\newcommand{\tr}{\mathrm{Tr}}
\newcommand{\respi}{\underset{w = \ic}{\mathrm{res}}}
\newcommand{\resmi}{\underset{w = -\ic}{\mathrm{res}}}

\newcommand{\qasq}{\quad \text{as} \quad}
\newcommand{\qandq}{\quad \text{and} \quad}

\newcommand{\ic}{\mathrm{i}}
\renewcommand{\O}{\mathcal{O}}

\begin{document}

\title[]{Long-time asymptotics of the autocorrelation function of the transverse Ising chain at the critical magnetic field Revisited}

\author{Noah Hout}



\author{Kenta Miyahara}



\author{Dustin Newland}



\author{Maxim Yattselev}

\address{Department of Mathematical Sciences, Indiana University Indianapolis, 402~North Blackford Street, Indianapolis, IN 46202}

\email{\href{mailto:nhout@iu.edu}{nhout@iu.edu},~~\href{mailto:kemiya@iu.edu}{kemiya@iu.edu},~~\href{mailto:dwnewlan@iu.edu}{dwnewlan@iu.edu},~~\href{mailto:maxyatts@iu.edu}{maxyatts@iu.edu}}

\begin{abstract}
Following \cite{DeiftZhou}, we analyze the long-time asymptotics of the autocorrelation function of the transverse Ising chain at the critical magnetic field (a special case of the spin-$\frac12$ XY model in a magnetic field) via the associated Riemann–Hilbert problem. 
We refine the original Deift-Zhou's result by determining the subleading growing term in the asymptotics.

\smallskip

\end{abstract}


\maketitle

\tableofcontents

\setlength{\parskip}{6pt}

\section{Introduction and Main Results}

Following the work of Deift-Zhou \cite{DeiftZhou}, we consider the 1D transverse Ising model at the critical transverse magnetic field. In this formulation, spins on a 1D lattice interact via nearest-neighbor exchange interaction along the $x$-axis and are subjected to a transverse magnetic field along the $z$-axis. The Hamiltonian for this model is given by
\begin{align}
    \mathcal{H} = -\frac{1}{2} \sum_{l \in \Z} \left( \sigma_l^x \sigma_{l+1}^x + \sigma_l^z \right),
\end{align}
where $\sigma_l^x$ and $\sigma_l^z$ are the spin Pauli matrices at the $l$-th site of the 1D lattice. The main object of study in \cite{DeiftZhou} is the autocorrelation function $\chi(t)$ between the first spin component $\sigma^x_0(t) = e^{-\ic \mathcal{H} t} \sigma^x_0(0) e^{\ic \mathcal{H} t}$ after time $t$ and the one at time $0$, i.e.,
\begin{align}
    \chi(t) = \frac{\tr\left( e^{-\beta \mathcal H} e^{-\ic \mathcal{H} t} \sigma^x_0(0) e^{\ic \mathcal{H} t} \sigma^x_0(0) \right)}{\tr \left(e^{-\beta \mathcal{H}} \right)},
\end{align}
where \( \beta \) is the inverse temperature.  More precisely, the long-time behavior of $\chi(t)$ is the main interest of Deift-Zhou. Prior to that, McCoy-Perk-Shrock \cite{MPS83} showed that $\chi(t)$ is given by the Fredholm determinant
\[
\chi(t) = e^{-t^2/2} \det \left(\mathcal{I} - \mathcal{K}_t \right),
\]
where $\mathcal{K}_t$ is a trace-class operator acting on $L^2((-1,1))$ with kernel
\[
K_t(z, w) = g(z) \frac{\sin (\ic t (z - w)) }{\pi (z - w)}, \quad z, w \in [-1,1].
\]
Here, we set $g(z) = \tanh ( \beta \sqrt{1 - z^2} ) > 0$ on $(-1,1)$, which fixes the branch of the square root.  As explained in \cite{IIKN90}, using the general theory of integrable operators, see \cite{IIKS90, KBI, Deiftintop} for example, one can show that this operator is associated with the following Riemann-Hilbert problem.

\begin{RHP} \label{rhp original}
Find a $2 \times 2$ matrix function $\Psi(z,t)$ such that
\begin{enumerate}
    \item $\Psi(z,t)$ is analytic for $z \in \C \setminus [-1,1]$;
    \item one-sided traces $\Psi_{\pm}(z,t)$ exist a.e. on \( (-1,1) \), are bounded there, and satisfy
    \begin{align*}
        \Psi_+(z,t) = \Psi_-(z,t) G_{\Psi}(z, t), \quad z \in (-1, 1),
    \end{align*}
    where $(-1, 1)$ is oriented from $-1$ to $1$ and 
    \begin{align} \label{defn of G Psi}
        G_{\Psi}(z, t) = \begin{pmatrix}
            1 + g(z) & -g(z)e^{2zt} \\
            g(z)e^{-2zt} & 1 - g(z) 
        \end{pmatrix}, \quad z \in (-1, 1);
    \end{align}
    \item it holds that \( \Psi(z,t) = I + z^{-1}\Psi_1(t) + \mathcal O(z^{-2})\) as \( z\to\infty\).
\end{enumerate}
\end{RHP}

Throughout the article, when we discuss the Riemann-Hilbert problems, we may omit explicit mention of the $t$-dependence of functions if no confusion is likely. Let
\begin{align}
\label{defn of sigma}
\sigma(t) := \frac{t^2}{2} + \ln \chi(t). 
\end{align}
The connection of the solution to RHP~\ref{rhp original} and $\chi(t)$ is given by
\begin{align}
\sigma^\prime(t) = -2[\Psi_1(t)]_{11}. \label{sigma and Psi1}
\end{align}
Thus, by solving RHP~\ref{rhp original} asymptotically for large $t$, one can find the long-time asymptotics of $\sigma'(t)$ from which Deift and Zhou obtained the following result.
\begin{DZthm}
    As $t \to \infty$, one has
    \begin{align} \label{DZ result eqn}
        \chi(t) = \exp \left[ \frac{t}{\pi} \int_{-1}^1 \ln| \tanh \beta \zeta| d \zeta + \O(\ln t) \right].
    \end{align}
\end{DZthm} 

Our goal is to find a more explicit expression of the error term $\O(\ln t)$ in \eqref{DZ result eqn}.
This is the problem posed by Percy Deift as one of the open problems at the conference\footnote{\href{https://sites.google.com/view/deift2025}{Universality, Nonlinearity, and Integrability}, May 2025, KIAS} in Seoul to celebrate his 80th birthday.

This type of problem has attracted significant attention in related settings. Indeed, let us recall that the 1D transverse Ising model at the critical transverse magnetic field is a particular case of the spin-$\frac12$ XY model in a magnetic field, see for example \cite{Henkel} and references therein. The Hamiltonian for this model is given by
\begin{align} \label{XY hamiltonian}
\mathcal{H}_{\text{XY}} = - \frac12 \sum_{l \in \Z} \left( (1 + \gamma) \sigma_l^x \sigma_{l+1}^x + (1 - \gamma) \sigma_l^y \sigma_{l+1}^y \right) - h \sum_{l \in \Z} \sigma_l^z,
\end{align}
where $\gamma$ is the anisotropic parameter and $h$ is the transverse magnetic field. 
Note that $\gamma$ interpolates between the Ising case ($\gamma = 1$) and the XXO Heisenberg chain ($\gamma = 0$). When we consider $\gamma = 1$ and $h = 1$ (the critical transverse magnetic field), then we indeed have
\[
\mathcal{H} = \frac12 \left.\mathcal{H}_{\text{XY}} \right|_{\gamma=1,h=1}.
\]
The other extremum, $\gamma=0$, has also been extensively studied. Analyzing the associated Riemann-Hilbert problem, Its-Izergin-Korepin-Slavnov \cite{IIKS93} found the asymptotics of the temperature correlation function in the spacelike direction and the timelike direction for a moderate magnetic field $0 \leq h < 1$ including the subleading growing term, see \cite[Equations (4) and (5)]{IIKS93}\footnote{The critical value of the magnetic field is \( h=2\) in \cite{IIKS93} due to a different normalization of the Hamiltonian.}. For the strong magnetic field $h > 1$, the timelike direction asymptotics was further studied by Jie, see \cite[Section 1.4]{Jie}.

A similar analysis was carried out by Its-Izergin-Korepin-Varzugin \cite{IIKV} to find the asymptotics of the two-point correlation function of an impenetrable Bose gas, where the role of an external magnetic field is played by the so-called chemical potential. Based on the sign of the chemical potential, there are three cases: (i) the case of negative chemical potential, (ii) the case of positive chemical potential in the spacelike direction, and (iii) the case of positive chemical potential in the timelike direction, see \cite[Section 8]{IIKV} for the asymptotics of the two-point correlation function for each case\footnote{One can make an analogy between the Bose gas paper \cite{IIKV} and the XXO Heisenberg chain papers \cite{Jie, IIKS93}; case (i) of \cite{IIKV} corresponds with \cite{Jie} and case (ii)-(iii) of \cite{IIKV} corresponds with \cite{IIKS93}.}. In \cite{IIKV}, the authors have found all the growing terms and even the constant terms. 

In the present paper, we study RHP~\ref{rhp original} of Deift and Zhou in more detail to obtain the subleading growing term of $\chi (t)$ as $t \to \infty$ from the analysis of the relevant small norm Riemann-Hilbert problem. The constant term has to be separately cared for, since the Riemann-Hilbert approach gives the asymptotics of $\sigma'(t)$ only, recall \eqref{defn of sigma} and \eqref{sigma and Psi1}, and this is our future project.
Our main result is formulated as the following theorem.

\begin{theorem}
\label{thm:main}
As $t \to \infty$, it holds for some constant \( C \) that
    \begin{align}
        \chi(t) = \exp \left[ \frac{t}{\pi} \int_{-1}^1 \ln| \tanh \beta \zeta| d \zeta + \frac{\ln^2(\tanh \beta) }{2\pi^2} \ln t + C + \O\big(t^{-1/4}\big) \right].
    \end{align}
\end{theorem}

For the above-described reasons, significant parts of the analysis in the present paper are based on that of \cite{DeiftZhou}. Hence, we repeatedly cite \cite{DeiftZhou} for proofs of certain statements. However, to point out the differences, we present all the steps of the calculations needed to show the subleading growing terms in the remaining sections.

Acknowledgments: The authors thank Alexander R. Its for helpful discussions and remarks. The second author was partially supported by a scholarship from the Japan Student Services Organization and the Hokushin Scholarship Foundation.


\section{Deformations of RHP~\ref{rhp original}}

In what follows, \( \sigma_1,\sigma_2,\sigma_3 \) are the Pauli matrices
\[
\sigma_1 = \begin{pmatrix} 0 & 1 \\ 1 & 0 \end{pmatrix}, \quad \sigma_2 = \begin{pmatrix} 0 & -\ic \\ \ic & 0 \end{pmatrix}, \quad \text{and} \quad \sigma_3 = \begin{pmatrix} 1 & 0 \\ 0 & -1 \end{pmatrix}.
\]

For our first step, we transfer RHP~\ref{rhp original} to the unit circle \( S^1=\{|w|=1\} \). To this end, let 
\[
\hat \Psi(w) := \Psi(z(w)), \quad z(w) := \frac{w - w^{-1}}{2\ic},
\]
which is analytic in $\C \setminus S^1$.  Since it must hold that $g(z(w)) \geq 0$ on $S^1$, we get that
\begin{align}
    g(z(w)) =  \begin{cases}
    h(w), & w \in S^1\cap\{\re(w)\geq0\}, \\
    -h(w), & w \in S^1\cap\{\re(w)\leq0\},
    \end{cases}
\end{align}
where
\begin{align} 
\label{defn of h}
    h(w) := \tanh \left( \beta \frac{w + w^{-1}}{2} \right).
\end{align}
As standard, we orient \( S^1 \) counter-clockwise. Notice that \( z(w) \) maps \( S^1\cap\{\re(w)\leq0\} \) into \( [-1,1] \), but oriented from \( 1 \) to \( -1 \). Thus, the jump matrix of \( \hat\Psi(w) \) satisfies
\begin{align}
G_{\hat \Psi} (w) & := \begin{Bmatrix} G_\Psi (z(w)), & w \in S^1\cap\{\re(w)\geq0\} \smallskip  \\
    G_\Psi (z(w))^{-1}, & w \in S^1\cap\{\re(w)\leq0\} \end{Bmatrix} \\
\label{defn of G Psi hat}
        & = \begin{pmatrix}
            1 + h(w) & -h(w) e^{-\ic t (w - w^{-1})} \\
            h(w) e^{\ic t (w - w^{-1})} & 1 - h(w)
        \end{pmatrix}, \quad w \in S^1.
\end{align}
Therefore, \( \hat \Psi  (w) \) solves the following Riemann-Hilbert problem.
\begin{RHP} 
\label{rhp Psi hat}
Find a $2 \times 2$ matrix function $\hat \Psi(w,t)$ such that
\begin{enumerate}
    \item $\hat \Psi(w,t)$ is analytic in $\overline\C \setminus S^1$;
    \item one-sided traces $\hat \Psi_{\pm}(w,t)$ exist a.e. on \( S^1 \), are bounded there, and satisfy
    \begin{align*}
        \hat \Psi_+(w,t) = \hat \Psi_-(w,t) G_{\hat \Psi}(w, t), \quad w \in S^1,
    \end{align*}
    where $G_{\hat \Psi}(w, t)$ is given by \eqref{defn of G Psi hat};
    \item it holds that \( \hat \Psi(w,t) = I + 2\ic w^{-1}\Psi_1(t) + \mathcal O(w^{-2})\) as \( w \to \infty\).
\end{enumerate}
\end{RHP}

Let us point out that RHP~\ref{rhp original} and RHP~\ref{rhp Psi hat} are simultaneously solvable and the solutions are related via $\hat \Psi(w) = \Psi(z(w))$. One direction is obvious by construction. In the other direction, given a solution of RHP~\ref{rhp Psi hat}, one needs to observe that it must satisfy
\[
\hat \Psi(w) = \hat \Psi(-1/w)
\]
since $G_{\hat \Psi}(w) = G_{\hat \Psi}(-1/w)$ as $h(w) = -h(-1/w)$, see \cite[Lemma 2.17]{DeiftZhou} (the ratio of these two functions is analytic across \( S^1 \) and is equal to \( I \) at infinity; the claim then follows from Liouville's theorem). This, in turn, allows one to define \( \Psi(z) \) out of \( \hat \Psi(w) \). In a similar vein, observe that \( G_{\hat \Psi}(w) = \sigma_1 G_{\hat \Psi}(-w) \sigma_1 \), which then induces the relation
\begin{align}
\label{Psi hat minus symmetry}
\hat \Psi(w) = \sigma_1 \hat \Psi(-w) \sigma_1, \quad w\in\overline\C\setminus S^1.
\end{align}

Next, we make the diagonal entries of jump matrix to be oscillatory by defining
\begin{align} 
\label{defn of Phi}
    \Phi(w) := \begin{cases}
        \hat \Psi (w) e^{\ic t w^{-1} \sigma_3}, & |w| > 1, \\
        \hat \Psi (w) e^{-\ic t w \sigma_3}, & |w| < 1.
    \end{cases}
\end{align}
Then, the jump matrix for \( \Phi(w) \) is given by
\begin{align}
    G_{\Phi}(w) & = e^{-\ic t w^{-1} \sigma_3} G_{\hat \Psi}(w) e^{-\ic t w \sigma_3} = \begin{pmatrix}
            (1 + h(w))e^{-t \varphi(w)} & -h(w) \\
            h(w) & (1 - h(w))e^{t \varphi(w)}
        \end{pmatrix},
\end{align}
where $\varphi(w)$ is purely imaginary on $S^1$ and is given by
\begin{align}
\label{defn of varphi}
    \varphi(w) := \ic (w + w^{-1}). 
\end{align}
Furthermore, RHP~\ref{rhp Psi hat}(3) becomes
\begin{align} \label{Phi normalization}
\begin{aligned}
    \Phi(w, t) &= \left( I + \frac{2\ic}{w}\Psi_1(t) + \mathcal O \left(w^{-2} \right) \right) \left( I + \frac{\ic t }{w}\sigma_3 + \O \left(w^{-2} \right)\right)\\
    &= I + w^{-1} \left( 2\ic \Psi_1(t) + \ic t \sigma_3 \right) + \O \left( w^{-2} \right), \quad w \to \infty.
\end{aligned}
\end{align}

Once one has an oscillatory jump matrix, the next standard step in non-linear steepest descent analysis is to open the lenses, see \eqref{Bruhat decomposition}. However, one technical issue arises from the fact that $h(w)$ vanishes on the imaginary axis including at  $\pm \ic \in S^1$, which complicates the desired factorization. This leads us to deform the jump contour $S^1$ away from $\pm \ic$. To this end, we introduce some notation. Let
\begin{align}
    \D_r(a) := \{w \in \C \,|\, |w - a| < r\} \qandq  \D := \D_1(0).
\end{align}
Also, we observe that $G_{\Phi}^{-1}(w)$ is analytic near $S^1$. Then, for $\rho \in (0,1)$ sufficiently small so that the disks $\D_{\rho}(\pm \ic)$ do not contain other zeros of \( h(w) \) besides \( \pm\ic \), we define $\tilde \Phi (w)$ by
\begin{align} 
\label{defn of Phi tilde}
    \tilde \Phi (w) := \begin{cases}
        \Phi(w) G_{\Phi}^{-1}(w), & w \in \D_{\rho}(\pm \ic) \cap \D, \\
        \Phi(w), & \text{elsewhere}.
    \end{cases}
\end{align}

\begin{figure}[ht!]
\centering
\includegraphics[width=5cm]{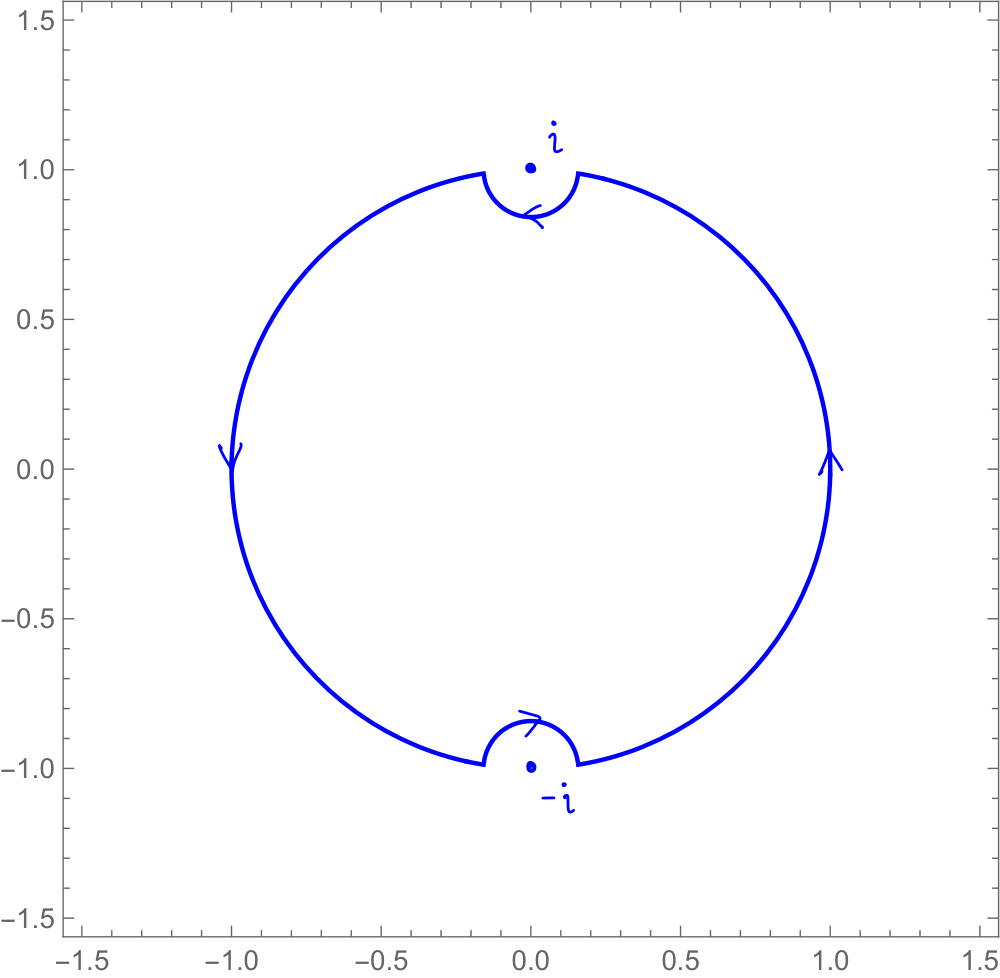}
\caption{The jump contour \( \Gamma_{gl} \).}
\label{Pic for Phi tilde RHP}
\end{figure}

Then, \( \tilde \Phi(w)\) solves the following Riemann-Hilbert problem.
\begin{RHP} 
\label{rhp Phi tilde}
Find a $2 \times 2$ matrix function $\tilde \Phi(w,t)$ such that
\begin{enumerate}
    \item $\tilde \Phi(w,t)$ is analytic in $\C \setminus \Gamma_{gl}$, where $\Gamma_{gl}$ is an oriented contour as on Figure \ref{Pic for Phi tilde RHP};
    \item one-sided traces $\tilde \Phi_{\pm}(w,t)$ exist a.e. on \( \Gamma_{gl} \), are bounded there, and satisfy
    \begin{align*}
        \tilde \Phi_+(w,t) = \tilde \Phi_-(w,t) G_{\tilde \Phi}(w, t), \quad w \in \Gamma_{gl},
    \end{align*}
    where
    \begin{align} \label{defn of G Phi tilde}
        G_{\tilde \Phi}(w, t) = \begin{pmatrix}
            (1 + h(w))e^{-t\varphi(w)} & -h(w) \\
            h(w) & (1 - h(w))e^{t \varphi(w)}
        \end{pmatrix};
    \end{align}
    \item it holds that \( \tilde \Phi(w,t) = I + w^{-1} (2\ic \Psi_1(t) + \ic t \sigma_3) + \mathcal O(w^{-2})\) as \( w \to \infty\).
\end{enumerate}
\end{RHP}
Clearly, RHP~\ref{rhp Psi hat} and RHP~\ref{rhp Phi tilde} are simultaneously solvable as the transformation connecting them is invertible.

\begin{figure}[ht!]
\centering
\includegraphics[width=7cm]{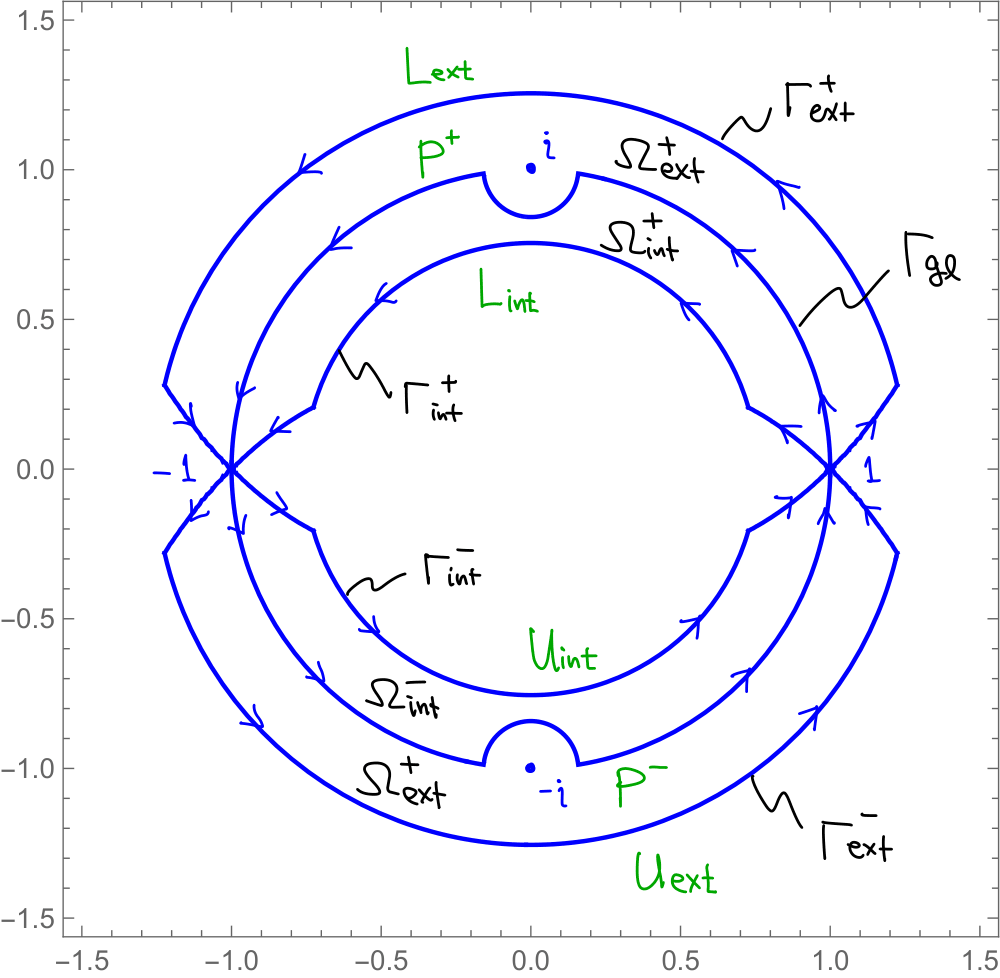}
\caption{The jump contour \( \Gamma:=\Gamma_{gl}\cup \Gamma^+_{ext}\cup \Gamma^+_{int}\cup \Gamma^-_{ext}\cup \Gamma^-_{int} \) and the jump matrices $G_{\hat \Phi}(w, t)$ for RHP~\ref{rhp Phi hat}. }
\label{Pic for Phi hat RHP}
\end{figure}

Next, we perform the Bruhat decomposition of \( G_{\tilde \Phi}(w) \). Namely, it holds that
\begin{align} 
\label{Bruhat decomposition}
    G_{\tilde \Phi}(w, t) = L_{ext}(w, t) P^+(w) L_{int}(w, t) = U_{ext}(w, t) P^-(w) U_{int}(w, t),
\end{align}
where
\begin{align} 
\label{defn of L1 P1 L2 U1 P2 U2}
\begin{aligned}
    L_{ext}(w, t) = \begin{pmatrix}
        1 & 0 \\
        \frac{h(w)-1}{h(w)}e^{t \varphi(w)} & 1
    \end{pmatrix}, & \quad L_{int}(w, t) = \begin{pmatrix}
        1 & 0 \\
        \frac{h(w)+1}{-h(w)} e^{-t\varphi(w)} & 1
    \end{pmatrix}, \\
    P^+(w) = \begin{pmatrix}
        0 & -h(w)\\
        \frac{1}{h(w)} & 0
    \end{pmatrix}, & \quad P^-(w) = \begin{pmatrix}
        0 & -\frac{1}{h(w)}\\
        h(w) & 0
    \end{pmatrix}, \\ 
    U_{ext}(w, t) = \begin{pmatrix}
        1 & \frac{h(w)+1}{h(w)}e^{-t \varphi(w)} \\
        0 & 1
    \end{pmatrix}, & \quad U_{int}(w, t) = \begin{pmatrix}
        1 & \frac{h(w)-1}{-h(w)} e^{t\varphi(w)} \\
        0 & 1
    \end{pmatrix}.
\end{aligned}
\end{align}
It readily follows from the explicit expression \eqref{defn of varphi} that
\begin{equation}
\label{sign chart}
\begin{cases}
\re(\varphi(w)) > 0, & \{|w|<1 ~\&~ \im(w)>0\}\cap \{|w|>1 ~\&~ \im(w)<0\}, \smallskip \\
\re(\varphi(w)) < 0, & \{|w|<1 ~\&~ \im(w)<0\}\cap \{|w|>1 ~\&~ \im(w)>0\}.
\end{cases}
\end{equation}
Hence, we use the first factorization in \eqref{Bruhat decomposition} around \( \Gamma_{gl}^+ := \Gamma_{gl} \cap \{\im(w)\geq0 \}\) and we use the second factorization in \eqref{Bruhat decomposition} around \( \Gamma_{gl}^- := \Gamma_{gl} \cap \{\im(w)\leq0 \} \). To this end, let the curves \( \Gamma_{int}^\pm,\Gamma_{ext}^\pm \) and the domains \( \Omega^\pm_{int}, \Omega^\pm_{ext} \) be as on Figure~\ref{Pic for Phi hat RHP}.
We assume that the curves \( \Gamma_{int}^\pm,\Gamma_{ext}^\pm \) are close enough to \( \Gamma_{gl} \) so that the only zeros of \( h(w) \) in \( \Omega^\pm_{int} \cup \Omega^\pm_{ext} \) are \( \pm\ic \). We also choose them so that \( \Gamma^-_{ext} = \{-w|w\in\Gamma_{ext}^+\} \) and \( \Gamma^-_{int} = \{-w|w\in\Gamma_{int}^+\} \). We set 
\begin{align}
\label{defn Phi hat}
\hat\Phi(w) := \tilde \Phi(w) 
\begin{cases}
L_{ext}(w,t), & w\in \Omega_{ext}^+, \\
L_{int}(w,t)^{-1}, & w\in \Omega_{int}^+, \\
U_{ext}(w,t), & w\in \Omega_{ext}^-, \\
U_{int}(w,t)^{-1}, & w\in \Omega_{int}^-, \\
I, & \text{otherwise}.
\end{cases}
\end{align}

Since $1/h(w)$ has simple poles at $\pm \ic$, each with residue $1/\beta$,  $L_{ext}(w, t)$ and $U_{ext}(w, t)$ have simple poles at $\ic$ and $-\ic$, respectively, and so does $\hat\Phi(w)$. Indeed, we have that
\[
L_{ext}(w) = \begin{pmatrix}
        1 & 0 \\
        \frac{h(w)-1}{h(w)} e^{t \varphi(w)} & 1
    \end{pmatrix} = \frac1{w - \ic} \begin{pmatrix}
        0 & 0\\
        -\frac{1}{\beta} & 0
    \end{pmatrix} + \begin{pmatrix} 1 & 0 \\ * & 1 \end{pmatrix} + \O(\big(w-\ic)\big)
\]
as \( w\to\ic \). Because \( \hat\Phi(w) = \tilde\Phi(w) L_{ext}(w) \) around \( \ic \), it holds that
\begin{equation}
\label{resudue i}
   \respi \hat\Phi(w) = \tilde\Phi(\ic) \begin{pmatrix}
        0 & 0\\
        -\frac{1}{\beta} & 0
    \end{pmatrix} = \lim_{w \to \ic} \tilde\Phi(w) \begin{pmatrix}
        0 & 0\\
        -\frac{1}{\beta} & 0
    \end{pmatrix} = \lim_{w \to \ic} \hat\Phi(w) \begin{pmatrix}
        0 & 0\\
        -\frac{1}{\beta} & 0
    \end{pmatrix} .    
\end{equation}
Similarly, at $w = -\ic$, we have
\begin{align}
\label{resudue -i}
   \resmi \hat\Phi(w) = \lim_{w \to -\ic} \hat\Phi(w) \begin{pmatrix}
        0 & \frac{1}{\beta}\\
        0 & 0
    \end{pmatrix}.
\end{align}
Altogether, $\hat\Phi(w)$ is the solution of the following Riemann-Hilbert problem.
\begin{RHP} 
\label{rhp Phi hat}
Find a $2 \times 2$ matrix function $\hat \Phi(w,t)$ such that
\begin{enumerate}
    \item $\hat \Phi(w,t)$ is analytic in $\C \setminus \left(\Gamma \cup \{\pm \ic\} \right)$, where $\Gamma$ is an oriented contour from Figure~\ref{Pic for Phi hat RHP};
    \item one-sided traces $\hat \Phi_{\pm}(w,t)$ exist a.e. on \( \Gamma \), are bounded there, and satisfy
    \begin{align*}
        \hat \Phi_+(w,t) = \hat \Phi_-(w,t) G_{\hat \Phi}(w, t), \quad w \in \Gamma,
    \end{align*}
    where the jump matrices $G_{\hat \Phi}(w, t)$ on $\Gamma$ are as in Figure \ref{Pic for Phi hat RHP};
    \item it holds that \( \hat \Phi(w,t) = I + w^{-1} (2 \ic \Psi_1(t) + \ic t \sigma_3) + \mathcal O(w^{-2})\) as \( w \to \infty\) by RHP~\ref{rhp Phi tilde}(3);
    \item $\hat \Phi(w, t)$ has simple poles at $w = \pm \ic$ with residues \eqref{resudue i} and \eqref{resudue -i}.
\end{enumerate}
\end{RHP}

Notice that the jump matrix \( G_{\hat \Phi}(w,t) \) decays exponentially fast to the identity matrix on \( \Gamma\setminus\Gamma_{gl} \) due to \eqref{sign chart}. 

As before, RHP~\ref{rhp Phi tilde} and RHP~\ref{rhp Phi hat} are simultaneously solvable. One direction is obvious. To go from  RHP~\ref{rhp Phi hat} to RHP~\ref{rhp Phi tilde} one needs to observe that
\begin{multline}
\hat\Phi(w) L_{ext}^{-1}(w) = \left( -\frac1\beta \begin{pmatrix} u & 0 \\ v & 0 \end{pmatrix} \frac1{w-\ic} + \begin{pmatrix} * & u \\ * & v \end{pmatrix} + \O(\big(w-\ic)\big)\right)\times \\
\left( \frac{1}{\beta} \begin{pmatrix}
        0 & 0\\
        1 & 0
    \end{pmatrix}\frac1{w - \ic} + \begin{pmatrix} 1 & 0 \\ * & 1 \end{pmatrix} + \O(\big(w-\ic)\big) \right) = \O(1)
\end{multline}
as \( w\to \ic \) by RHP~\ref{rhp Phi hat}(4) for some values \( u,v \) (these are the values of the second column of \( \hat\Phi(w)\) at \( \ic \)). As an analogous computation holds around \( -\ic \), \eqref{defn Phi hat} is indeed reversible.

The jump matrix \( G_{\hat\Phi}(w)\) on \( \Gamma_{gl} \) is independent of \( t \), but is a function of \( w \). In the next step, we turn it into a constant matrix. To this end, we set
\begin{align} 
\label{defn of Y}
    Y(w) := \begin{cases}
        \hat\Phi(w) \delta(w)^{\sigma_3}, & w\in\mathrm{ext}(\Gamma_{gl}),\\
        \hat\Phi(w) \delta(w)^{-\sigma_3}, & w\in\mathrm{int}(\Gamma_{gl}),
    \end{cases}
\end{align}
where \( \mathrm{int}(\Gamma_{gl}) \) and \( \mathrm{ext}(\Gamma_{gl}) \) are the interior and exterior domains of \( \Gamma_{gl} \), respectively, and $\delta(w)$ is the solution of the following scalar Riemann-Hilbert problem.

\begin{RHP} \label{rhp delta}
Find a scalar function $\delta (w)$ such that
\begin{enumerate}
    \item $\delta (w)$ is analytic in $\C \setminus \Gamma_{gl}$;
    \item one-sided traces $\delta _{\pm}(w)$ exist a.e. on \( \Gamma_{gl} \), are bounded there, and satisfy
    \begin{align*}
        \delta_+(w) = \delta_-(w)\begin{cases}
            h(w)^{-1}, & w \in \Gamma_{gl}^+,\\
            h(w), & w \in \Gamma_{gl}^-;
        \end{cases}
    \end{align*}
    \item it holds that \( \delta (w) \to 1 \) as \( w \to \infty\).
\end{enumerate}
\end{RHP}

By the Plemelj-Sokhotski formula, the solution of RHP~\ref{rhp delta} is given by
\begin{align} 
\label{PS formula for delta}
    \delta(w) = \exp \left\{ \frac{1}{2\pi \ic} \int_{\Gamma_{gl}^+} \frac{-\log h(s)}{s - w} ds + \frac{1}{2\pi \ic} \int_{\Gamma_{gl}^-} \frac{\log h(s)}{s - w} ds \right\},
\end{align}
where we take
\begin{align} 
\label{principal branch of ln}
    \log h(s) = \begin{cases}
    \ln |h(s)|, & s \in S^{1} \cap \{\re (s) > 0\}, \\
    \ln |h(s)| - \ic \pi, & s \in S^{1} \cap \{\re (s) < 0~\&~\im (s) > 0\}, \\
    \ln |h(s)| + \ic \pi, & s \in S^{1} \cap \{\re (s) < 0~\&~\im (s) < 0\},
    \end{cases}
\end{align}
which ensures boundedness of \( \delta(w) \) around \( \pm1 \) (discussion following \eqref{delta of w near w=1} further below implies that \( |\delta(w)| \) is bounded above and away from zero in a neighborhood of \( 1 \), boundedness around \( -1 \) then follows from \eqref{delta(w) and delta(-w)}). Since $|h(-s)|=|h(s)|$, one can use \eqref{PS formula for delta} and \eqref{principal branch of ln} to check that
\begin{align} 
\label{delta(w) and delta(-w)}
\delta(w) \delta(-w) = \begin{dcases}
    1, & w\in\mathrm{ext}(\Gamma_{gl}),\\
    -1, & w\in\mathrm{int}(\Gamma_{gl}).
\end{dcases}
\end{align}
Moreover, it was shown in \cite[Equations~(3.19) and (3.21)]{DeiftZhou} that
\begin{align}
\label{delta pm i}
\delta(\ic) = \sqrt{2 \beta} \quad \text{and} \quad \delta(-\ic) = 1/\sqrt{2 \beta}.
\end{align}
Finally, \(\ \delta(w) = 1 + w^{-1}\delta_1 + \O(w^{-2}) \) as \( w \to \infty \), where, see \cite[Equation~(3.30)]{DeiftZhou}, 
\begin{align} 
\label{defn of delta 1}
    \delta_1 = \frac{\ic}{\pi} \int_{-1}^1 \ln |\tanh \beta \zeta| d\zeta + \ic.
\end{align}

Coming back to the matrix \( Y(w) \), it readily follows from \eqref{defn of Y} and RHP~\ref{rhp delta}(2) that $Y(w)$ admits the jump condition $Y_+(w) = Y_-(w) G_Y(w)$, where
\begin{align} 
\label{defn of G Y}
    G_Y(w) = \begin{cases}
        \delta(w)^{\sigma_3} G_{\hat{\Phi}}(w) \delta(w)^{-\sigma_3}, & w \in \Gamma^+_{int}\cup \Gamma^-_{int},\\
        -\ic \sigma_2, & w \in \Gamma_{gl},\\
        \delta(w)^{-\sigma_3} G_{\hat{\Phi}}(w) \delta(w)^{\sigma_3}, & w \in \Gamma^+_{ext}\cup \Gamma^-_{ext}.
    \end{cases}
\end{align}
Furthermore, it follows from \eqref{defn of Y} and RHP~\ref{rhp Phi hat}(3) that
\begin{align} 
\label{Y(w) near infty}
    Y(w, t) = I + w^{-1} \big( 2 \ic \Psi_1(t) + (\ic t+\delta_1) \sigma_3 \big) + \O\big(w^{-2}\big)
\end{align}
as \( w\to\infty \).
Next, we get from \eqref{defn of Y}, RHP~\ref{rhp Phi hat}(4), and \eqref{delta pm i} that
\begin{align}
    \respi Y (w) &= \respi \hat\Phi (w) \delta(\ic)^{\sigma_3} = \lim_{w \to \ic}  \hat{\Phi}(w) \begin{pmatrix}
        0 & 0\\
        -\frac{1}{\beta} & 0
    \end{pmatrix} \delta(\ic)^{\sigma_3}\\
    &= \lim_{w \to \ic} Y(w) \delta(\ic)^{-\sigma_3} \begin{pmatrix}
        0 & 0\\
        -\frac{1}{\beta} & 0
    \end{pmatrix} \delta(\ic)^{\sigma_3}
    = \lim_{w \to \ic} Y(w) \begin{pmatrix}
        0 & 0\\
        -2 & 0
    \end{pmatrix}.
\end{align}
Clearly, the residue at $w = -\ic$ can be computed similarly. Thus, $Y(w)$ is the solution of the following Riemann-Hilbert problem.
\begin{RHP} 
\label{rhp Y}
Find a $2 \times 2$ matrix function $Y(w,t)$ such that
\begin{enumerate}
    \item $Y(w,t)$ is analytic in $\C \setminus \left(\Gamma \cup \{\pm \ic\} \right)$;
    \item one-sided traces $Y_{\pm}(w)$ exist a.e. on \( \Gamma \), are bounded there, and satisfy
    \begin{align*}
        Y_+(w,t) = Y_-(w,t)G_Y(w,t), \quad w \in \Gamma,
    \end{align*}
    where $G_Y(w, t)$ is given by \eqref{defn of G Y};
    \item it holds that \( Y(w,t) = I + w^{-1} Y_1(t) + \mathcal O(w^{-2})\) as \( w \to \infty\), where
    \begin{align}
        Y_1(t) = 2 \ic \Psi_1(t) + (\ic t +\delta_1) \sigma_3;
    \end{align}
    \item $Y(w,t)$ has simple poles at $w = \pm \ic$ with residues
    \begin{align}
    \respi Y(w) = \lim_{w \to \ic}  Y(w) \begin{pmatrix}
        0 & 0\\
        -2 & 0
    \end{pmatrix}  \quad \text{and} \quad
    \resmi Y(w) = \lim_{w \to -\ic} Y(w) \begin{pmatrix}
        0 & 2\\
        0 & 0
    \end{pmatrix} .
    \end{align}
\end{enumerate}
\end{RHP}

Since the transformation \eqref{defn of G Y} is clearly invertible, RHP~\ref{rhp Phi hat} and RHP~\ref{rhp Y} are simultaneously solvable.  For future use, observe that
\begin{align}
\label{GY symmetry}
    G_{Y}(w) = \sigma_1 G_{Y}(-w) \sigma_1 \begin{dcases}
        1, & w \in \Gamma \setminus \Gamma_{gl},\\
        -1, & w \in \Gamma_{gl},
    \end{dcases}
\end{align}
by \eqref{defn of L1 P1 L2 U1 P2 U2}, \eqref{delta(w) and delta(-w)}, and \eqref{defn of G Y}. It is not immediate that \eqref{GY symmetry} implies the corresponding symmetry of \( Y(w) \) due to the presence of poles in RHP~\ref{rhp Y}. However, we have already established that a solution RHP~\ref{rhp Y} produces a solution of RHP~\ref{rhp Psi hat} by reversing \eqref{defn of Y}, \eqref{defn Phi hat}, \eqref{defn of Phi tilde}, and \eqref{defn of Phi}. Then, \eqref{Psi hat minus symmetry} and \eqref{delta(w) and delta(-w)} do give
\begin{align}
\label{sigma1 symmetry of Y}
    Y(w) = \sigma_1 Y(-w) \sigma_1 \begin{dcases}
        1, & w\in\mathrm{ext}(\Gamma_{gl}), \\
        -1, & w\in\mathrm{int}(\Gamma_{gl}).
    \end{dcases}
\end{align}
It also follows from \eqref{sigma and Psi1} and RHP~\ref{rhp Y}(3) that
\begin{align} 
\label{sigma and Y1}
    \sigma'(t) = \ic \left[ Y_1(t) \right]_{11} - \ic \delta_1 + t.
\end{align}


\section{Global Parametrix} \label{global parametrix section}

As we mentioned before, the jump matrices in RHP~\ref{rhp Phi hat} tend to the identity matrix as $t \to \infty$ on $\Gamma \setminus \Gamma_{gl}$. Moreover, we shall show later that \( \delta(w) \) is bounded there. Hence, the leading order behavior of \( Y(w) \) is determined by its jump on \( \Gamma_{gl} \). Thus, we consider the following global parametrix  Riemann-Hilbert problem.
\begin{RHP} \label{rhp P gl}
Find a $2 \times 2$ matrix function $P^{(gl)}(w)$ such that
\begin{enumerate}
    \item $P^{(gl)}(w)$ is analytic in $\overline \C \setminus \left(\Gamma_{gl} \cup \{\pm \ic\} \right)$;
    \item one-sided traces $P^{(gl)}_{\pm}(w)$ exist a.e. on \( \Gamma_{gl} \), are bounded there, and satisfy
    \begin{align*}
        P^{(gl)}_+(w) = P^{(gl)}_-(w) \begin{pmatrix}
            0 & -1\\
            1 & 0
        \end{pmatrix}, \quad w \in \Gamma_{gl};
    \end{align*}
    \item it holds that \( P^{(gl)} (w) = I + w^{-1} P^{(gl)}_1 + \O(w^{-2}) \) as \( w \to \infty\);
    \item $P^{(gl)} (w)$ has simple poles at $w = \pm \ic$ with residues
    \begin{align}
    \respi P^{(gl)} (w) &= \lim_{w \to \ic} P^{(gl)} (w) \begin{pmatrix}
        0 & 0\\
        -2 & 0
    \end{pmatrix}, \\
    \resmi P^{(gl)} (w) &= \lim_{w \to -\ic} P^{(gl)} (w) \begin{pmatrix}
        0 & 2\\
        0 & 0
    \end{pmatrix}.
    \end{align}
\end{enumerate}
\end{RHP}

An explicit solution of RHP~\ref{rhp P gl} is given by
\begin{align}
\label{P global}
P^{(gl)} (w) := A(w) B, \quad P^{(gl)}_1 = \begin{pmatrix}
    \ic & 1 \\
    -1 & - \ic
\end{pmatrix},
\end{align}
where
\begin{equation} 
\label{defn of P gl}
        A(w) := \begin{pmatrix}
            \displaystyle \frac{w}{w - \ic} & \displaystyle \frac{1}{w + \ic} \smallskip \\
            \displaystyle  \frac{-1}{w - \ic} & \displaystyle \frac{w}{w + \ic}
        \end{pmatrix} \qandq B := \begin{cases}
        I, & w\in\mathrm{ext}(\Gamma_{gl}),\\
        -\ic \sigma_2, & w\in\mathrm{int}(\Gamma_{gl}).
    \end{cases}
\end{equation}
One can readily check that
\begin{align}
\label{sigma1 symmetry of Pgl}
    P^{(gl)}(w) = \sigma_1 P^{(gl)}(-w) \sigma_1 \begin{dcases}
        1, & w\in\mathrm{ext}(\Gamma_{gl}),\\
        -1, & w\in\mathrm{int}(\Gamma_{gl}).
    \end{dcases}
\end{align}

\section{Approximate Local Parametrices} \label{local parametrix section}

Since the jump matrices forming \( G_{Y}(w, t) \) are not uniformly close to the identity matrix in a small disk around $w = \pm1$, we need to solve RHP~\ref{rhp Y} locally in \( \mathbb{D}_\epsilon(\pm1) \) for some $\epsilon>0$ fixed. There are many such solutions, as any of them can be multiplied by a holomorphic matrix function on the left. We seek ones that match well the global parametrix on \( \partial \mathbb{D}_\epsilon(\pm1) \).


\subsection{Local Deformation of RHP~\ref{rhp Y}}

To construct approximate local solution of RHP~\ref{rhp Y} in \( \D_\epsilon(1) \), we first remove the jump of \( Y(w) \) on $\Gamma_{gl} \cap \D_{\epsilon}(1)$ by setting
\begin{align} 
\label{defn of Y tilde near 1}
    \tilde Y (w, t) = Y(w, t) B^{-1},
\end{align}
where $B$ was defined in \eqref{defn of P gl}. For labeling reasons it will convenient for us to write
\[
\begin{cases}
\Gamma_0 = - \Gamma_{int}^- \cap \D_\epsilon(1), & \Gamma_2 = \Gamma_{ext}^+ \cap \D_\epsilon(1), \\
\Gamma_1 = - \Gamma_{ext}^- \cap \D_\epsilon(1), & \Gamma_3 =  \Gamma_{int}^+ \cap \D_\epsilon(1)
\end{cases}
\]
\begin{figure}[ht!]
    \centering
    \includegraphics[width=5cm]{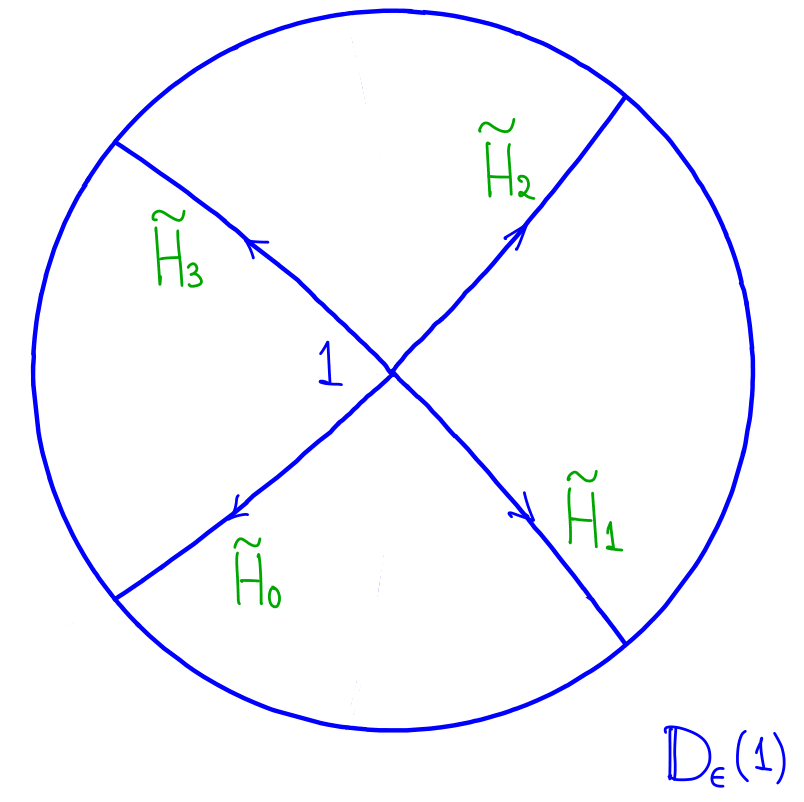}
    \caption{Arcs \( \Gamma_i \) and jump matrices \( \tilde H_i(w,t) \) for $\tilde Y(w,t)$.}
    \label{Y tilde jump matrices}
\end{figure}
(we flipped orientations of \( \Gamma^-_{int} \) and \( \Gamma^-_{ext} \)). Then, according to \eqref{defn of L1 P1 L2 U1 P2 U2} and \eqref{defn of G Y}, $\tilde Y(w, t)$ admits jump conditions as shown in Figure~\ref{Y tilde jump matrices}, where
\begin{align}
\label{defn of H0 tilde - H3 tilde}
\begin{aligned}
    \tilde H_0(w,t) = \begin{pmatrix}
        1 & 0\\
        \delta(w)^2 \frac{1-h(w)}{h(w)} e^{t\varphi(w)} & 1
    \end{pmatrix}, 
    & \;\; \tilde H_1(w,t) = \begin{pmatrix}
        1 & \frac{-1}{\delta(w)^2} \frac{h(w) + 1}{h(w)} e^{-t\varphi(w)}\\
        0 & 1
    \end{pmatrix},\\
    \tilde H_2(w,t) = \begin{pmatrix}
        1 & 0\\
        \delta(w)^2 \frac{h(w) - 1}{h(w)} e^{t\varphi(w)} & 1
    \end{pmatrix}, 
    & \;\; \tilde H_3(w,t) = \begin{pmatrix}
        1 & \frac{1}{\delta(w)^2} \frac{h(w) + 1}{h(w)} e^{-t\varphi(w)}\\
        0 & 1
    \end{pmatrix}.
\end{aligned}
\end{align}

To further transform $\tilde Y(w, t)$ near $w = 1$, we study the behavior of $\delta(w)$ there. To this end, we define $K_{1,2,3}$ to be the subregion of  $\D_\epsilon(1)$ as indicated on Figure~\ref{Pictures of K^{(i)}}, and set
\begin{align} 
\label{defn of nu and h hat}
    \nu := \frac{1}{\pi \ic} \ln h(1) \in \ic \R, \quad \hat{h}(s) := \begin{cases}
        h(1)h(s)^{-1}, & s \in \Gamma_{gl}^+,\\
        h(s)h(1)^{-1}, & s \in \Gamma_{gl}^-.
    \end{cases}
\end{align}

\begin{figure}[ht!]
    \centering
    \begin{subfigure}{.33\textwidth}
    \includegraphics[width=0.8\linewidth]{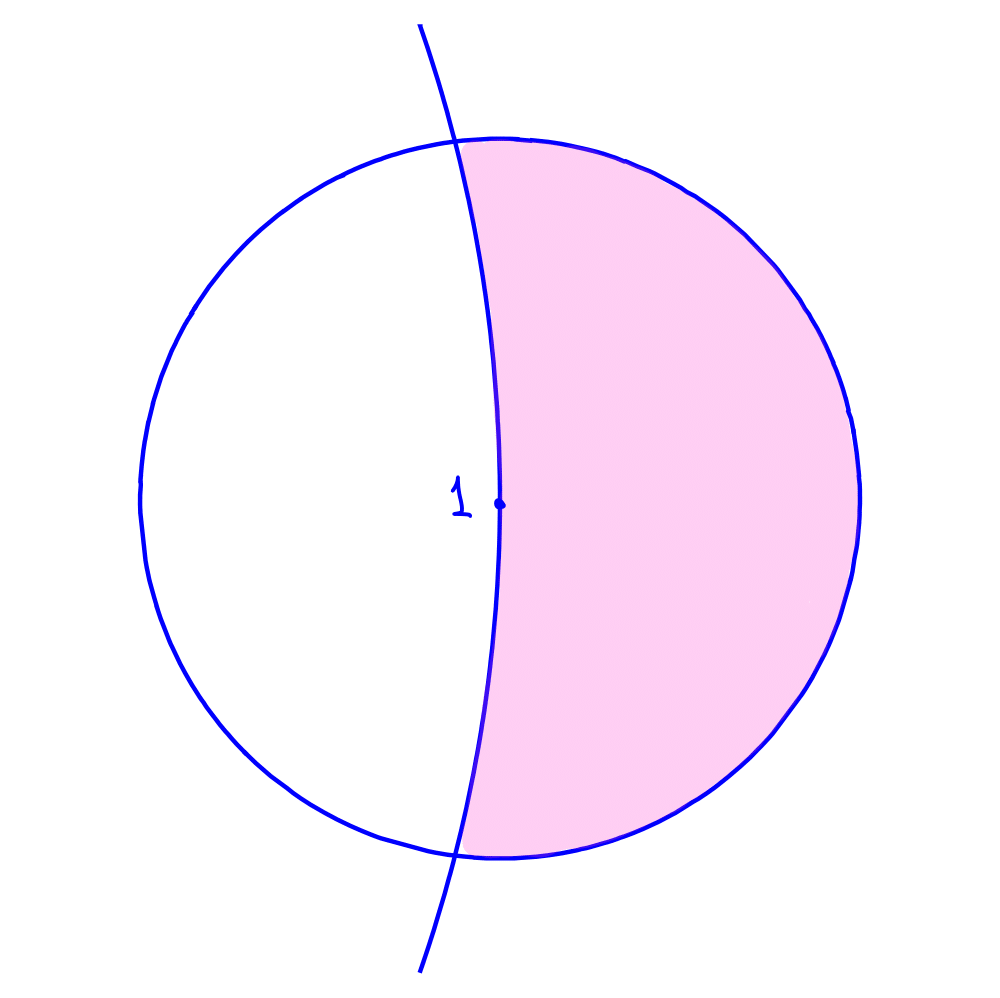}
    \caption{$K_1$}
    \label{Picture of K^{(1)}}
    \end{subfigure}%
    \begin{subfigure}{.33\textwidth}
    \includegraphics[width=0.8\linewidth]{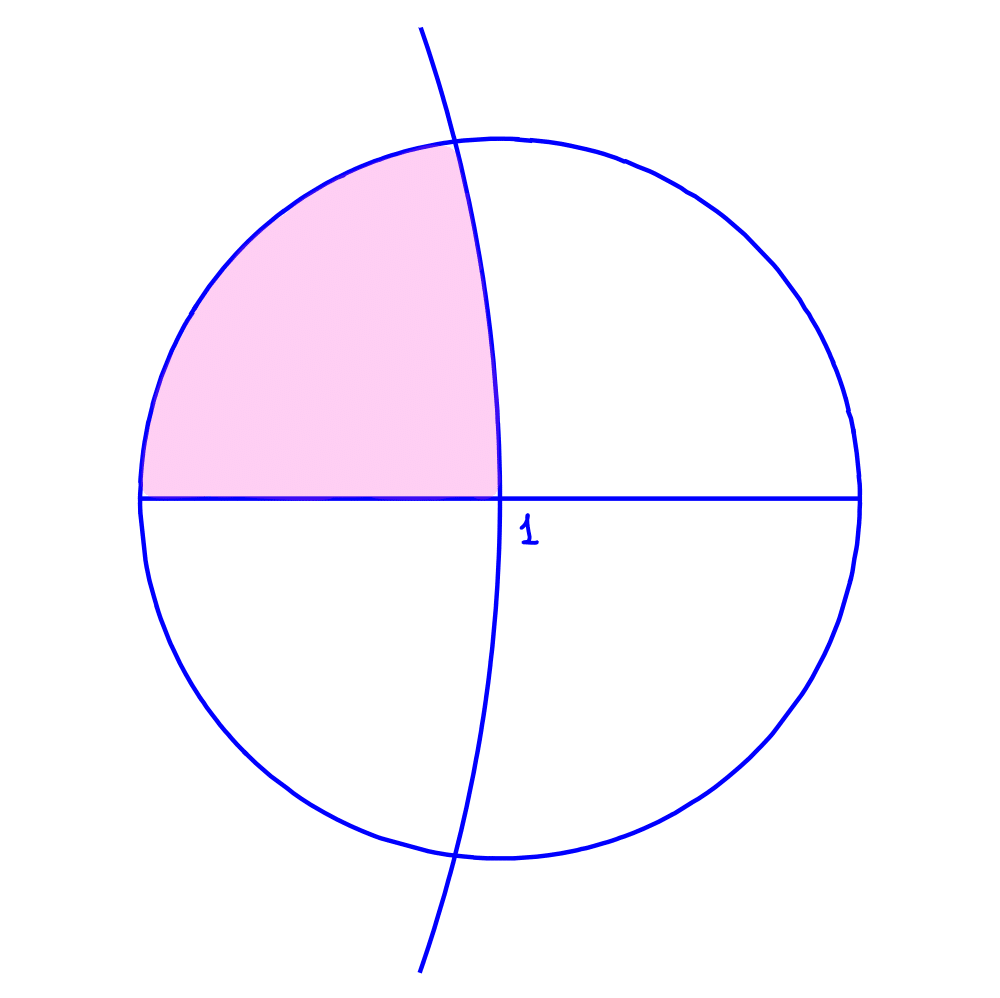}
    \caption{$K_2$}
    \label{Picture of K^{(2)}}
    \end{subfigure}
    \begin{subfigure}{.33\textwidth}
    \includegraphics[width=0.8\linewidth]{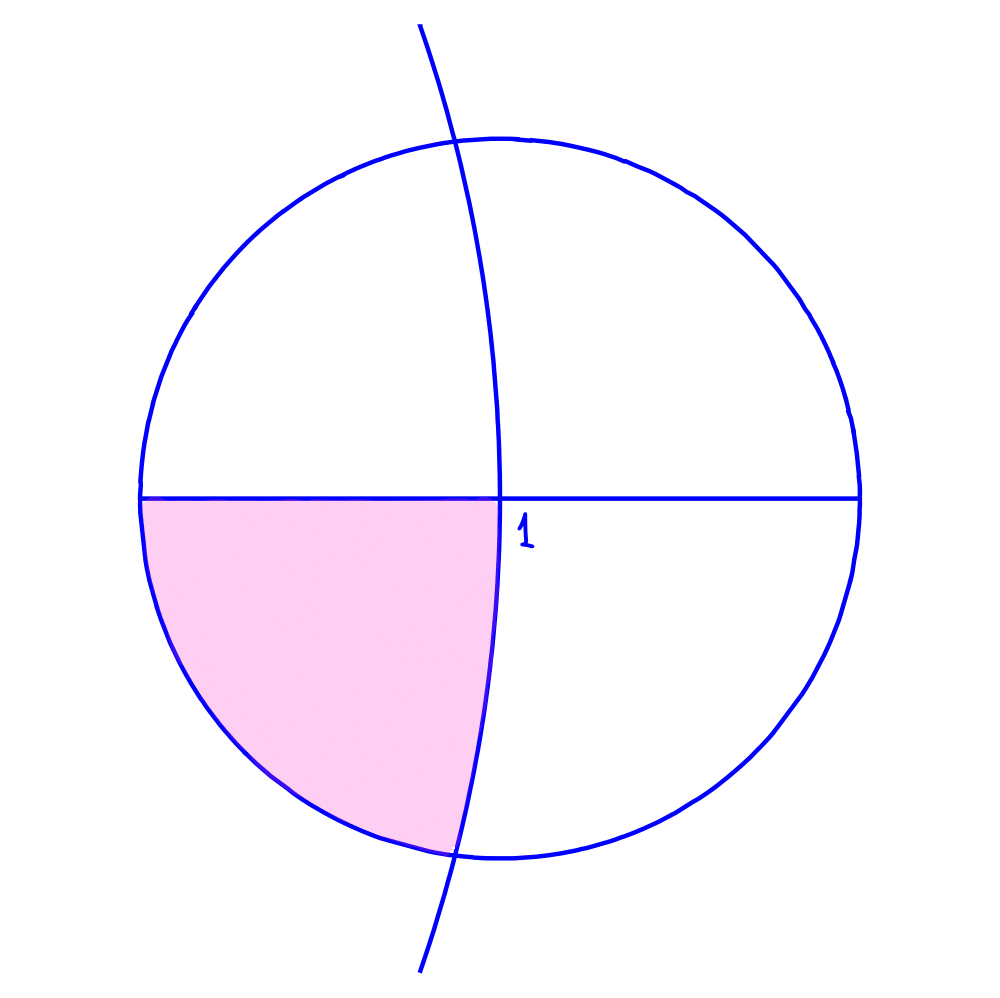}
    \caption{$K_3$}
    \label{Picture of K^{(3)}}
    \end{subfigure}
    \caption{Regions $K_{1,2,3}$.}
    \label{Pictures of K^{(i)}}
\end{figure}
Then, \eqref{PS formula for delta} can be written as
\begin{align}
\delta(w) &= \exp\left\{ \frac{1}{2\pi \ic} \int_{\Gamma_{gl}} \frac{\log \hat h(s)}{s - w} ds \right\} \exp \left\{ -\frac{\nu}{2} \int_{\Gamma_{gl}^+} \frac{ds}{s - w} + \frac{\nu}{2} \int_{\Gamma_{gl}^-} \frac{ds}{s - w} \right\} \\
&= \exp\left\{ \frac{1}{2\pi \ic} \int_{\Gamma_{gl}} \frac{\log \hat h(s)}{s - w} ds \right\} \left( \frac{w-1}{w+1}\right)^{\nu/2}_{\Gamma^+_{gl}}\left( \frac{w-1}{w+1}\right)^{\nu/2}_{\Gamma^-_{gl}},
\end{align}
where the power functions are positive for \( w>1 \) and subindex \( \Gamma_{gl}^\pm \) indicates their respective branch-cuts. This means that
\begin{align}
\left( \frac{w-1}{w+1}\right)^{\nu/2}_{\Gamma^\pm_{gl}} = \left( \frac{w-1}{w+1}\right)^{\nu/2} e^{\mp\pi\ic\nu}, \quad w\in \{|w|<1\} \cap \{\pm\im(w)>0\},
\end{align}
where the root on the right-hand side of the above equality is principal. Thus,
\begin{align} 
\label{delta of w near w=1}
\delta(w) = (w - 1)^{\nu} \hat \delta(w) \begin{cases}
    1, & w \in K_1,\\
    e^{-\pi \ic \nu}, & w \in K_2,\\
    e^{\pi \ic \nu}, & w \in K_3,
    \end{cases}
\end{align}
where the root is principal and \( \hat \delta(w) \) is an analytic function \( \D_\epsilon(1)\setminus \Gamma_{gl}\) given by
\begin{align} 
\label{delta hat of w}
    \hat \delta(w) = \left( w+1 \right)^{-\nu}\exp\left\{ \frac{1}{2\pi \ic} \int_{\Gamma_{gl}} \frac{\log \hat h(s)}{s - w} ds \right\}.
\end{align}
Since \( \nu \) is purely imaginary, \( (w-1)^\nu \) is a bounded function around \( 1 \).  We claim that $\hat\delta(w)$ is bounded there as well. We get from \eqref{defn of nu and h hat} that
\[
(\log \hat h(s))^\prime = \mp h^\prime(s) h^{-1}(s), \quad s\in \Gamma^\pm_{gl}\setminus\{1\}.
\]
Since \( h^\prime(1) =0 \), \( \log \hat h(s) \) is not only continuous on \( \Gamma_{gl}\cap \D_\epsilon(1) \) (it is equal to 0 at 1), but is also continuously differentiable. Hence, \( \hat \delta(w) \) extends continuously to  \( \Gamma_{gl}\cap \D_\epsilon(1) \). In fact, these extensions (from the left- and right-hand sides of \( \Gamma_{gl} \)) are continuously differentiable. Indeed, by \cite[Section~4.4]{Gakhov} we need to show that the derivative extends smoothly to \( \Gamma_{gl}\cap \D_\epsilon(1) \). It holds that
\begin{align} 
\hat \delta^\prime (w) = \left( -\frac\nu{w + 1} + \frac{1}{2\pi\ic} \int_{\Gamma_{gl}} \frac{\log \hat h(s)}{(s - w)^2} ds \right) \hat \delta(w).
\end{align}
Using integration by parts and \eqref{PS formula for delta} gives
\begin{align}
\label{delta'(w)}
\hat \delta^\prime (w) = \left( -\frac\nu{w + 1} + \frac{1}{2\pi\ic} \int_{\Gamma_{gl}} \frac{(\log \hat h(s))^\prime}{s - w} ds \right) \hat \delta(w).
\end{align}
One can readily check that \( \log \hat h(s) \) is, in fact, twice continuously differentiable on \( (\Gamma_{gl}\setminus\{1\})\cap \D_\epsilon(1) \) except for a  jump discontinuity of the second derivative at \( 1 \). Hence, \( (\log \hat h(s))^\prime \) is Lipschitz continuous and respectively  \( \hat \delta^\prime (w) \) extends smoothly to \( \Gamma_{gl} \cap \D_\epsilon(1) \) (as a H\"older continuous function of any index \(<1\)). Since \( (\log \hat h(s))^\prime \) vanishes at \( 1 \), this again implies that \( \hat \delta^\prime(1) \) is well defined.

\begin{figure}[ht!]
\centering
\includegraphics[width=5cm]{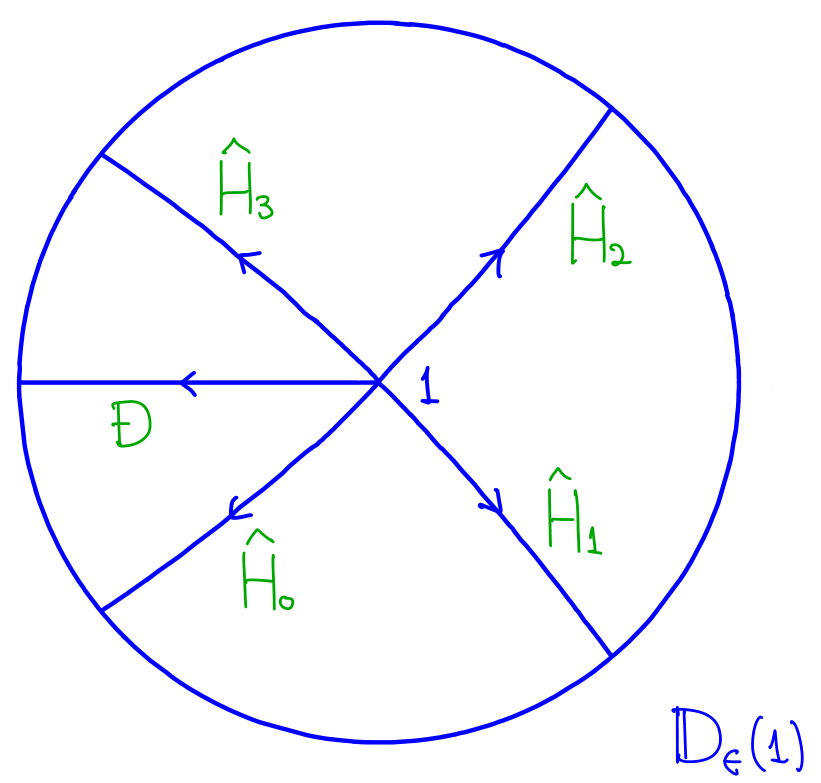}
\caption{The jump contour and the jump matrices for $\hat{Y}(w, t)$.}
\label{contour and matrix for Y hat}
\end{figure}

Going back to the matrix \( \tilde Y(w, t) \), let
\begin{align} 
\label{defn of Y hat near 1}
    \hat{Y}(w, t) = \tilde Y(w, t) (w - 1)^{-\nu \sigma_3}, \quad w \in \D_{\epsilon}(1),
\end{align}
where the branch of the power function is principal. This transformation creates a jump across $(-1,1) \cap \D_{\epsilon}(1)$ while the jump matrices for $\tilde Y(w, t)$ are conjugated by $(w - 1)^{\nu \sigma_3}$. More precisely, the jump matrices for $\hat Y(w, t)$ are shown on Figure~\ref{contour and matrix for Y hat}, where
\begin{align}
\begin{aligned}
    \hat H_0(w, t) &= \begin{pmatrix}
        1 & 0\\
        \hat \delta(w)^2 e^{2\pi \ic \nu} \frac{1-h(w)}{h(w)} e^{t\varphi(w)} & 1
    \end{pmatrix}, \quad w\in\Gamma_0, \\ 
    \hat H_1(w, t) &= \begin{pmatrix}
        1 & -\frac{1}{\hat \delta(w)^2} \frac{h(w) + 1}{h(w)} e^{-t\varphi(w)}\\
        0 & 1
    \end{pmatrix},  \quad w\in\Gamma_1, \\
    \hat H_2(w,t) &= \begin{pmatrix}
        1 & 0\\
        \hat \delta(w)^2 \frac{h(w) - 1}{h(w)} e^{t\varphi(w)} & 1
    \end{pmatrix},  \quad w\in\Gamma_2, \\ 
    \hat H_3(w,t) &= \begin{pmatrix}
        1 & \frac{1}{\hat \delta(w)^2} e^{2\pi \ic \nu} \frac{h(w) + 1}{h(w)} e^{-t\varphi(w)}\\
        0 & 1
    \end{pmatrix},  \quad w\in\Gamma_3, \\
    D & =e^{2\pi \ic \nu \sigma_3},  \quad w\in(1-\epsilon,1),
\end{aligned} \label{defn of H0hat-H3hat D}
\end{align}
by \eqref{defn of H0 tilde - H3 tilde}, \eqref{delta of w near w=1}, and \eqref{defn of Y hat near 1} (notice that we oriented the segment \( (1-\epsilon,1) \) away from 1).


\subsection{Parabolic Cylinder Model Riemann-Hilbert Problem}

It readily follows from \eqref{defn of varphi} that
\begin{align}
    \varphi(w) = 2\ic + \ic \frac{(w - 1)^2}w,
    \label{varphi near 1}    
\end{align}
where the second summand can be interpreted as a square of a conformal map, see \eqref{zeta <-> w}. Then, by replacing \( \hat \delta(w), h(w) \) by \( \hat \delta(1), h(1) \), and \( t\varphi(w) \) by \( 2\ic t - \zeta^2/2 \), we formally arrive at the following model Riemann-Hilbert problem.

\begin{figure}[htbp]
    \centering
    \includegraphics[width=6cm]{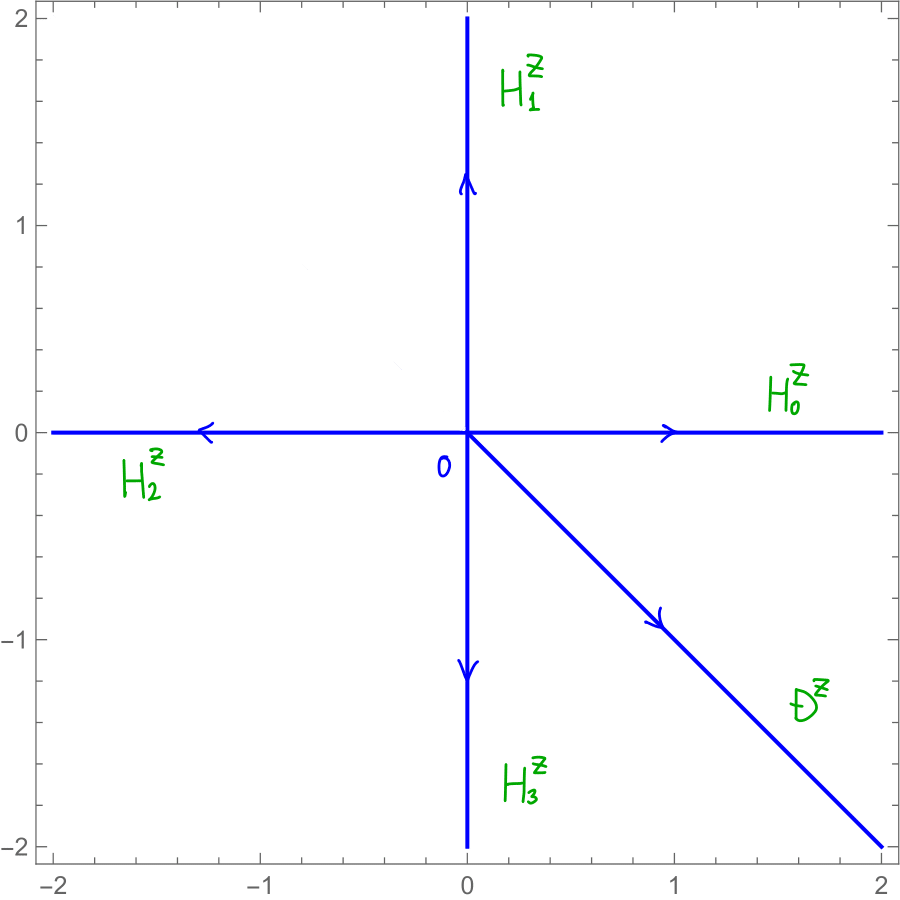}
    \caption{The contour $\Gamma_Z$, which consists of coordinate axes and the ray \( \arg(\zeta) = -\pi /4 \), and the jump matrices $G_{Z}(\zeta)$ for RHP~\ref{parabolic cylinder rhp}.}
    \label{pic for the Z problem}
\end{figure}

\begin{RHP}\label{parabolic cylinder rhp}
Find a $2 \times 2$ matrix function $Z(\zeta)$ such that
\begin{enumerate}
    \item $Z(\zeta)$ is analytic for $\zeta \in \C \setminus \Gamma_{Z}$, where $\Gamma_Z$ is depicted in Figure~\ref{pic for the Z problem};
    \item $Z(\zeta)$ has continuous traces on $\Gamma_Z \setminus \{0\}$ that satisfy
    \begin{align*}
        Z_+(\zeta) = Z_-(\zeta) G_{Z}(\zeta), \quad \zeta \in \Gamma_{Z},
    \end{align*}
    where the jump matrices comprising $G_{Z}(\zeta)$ are as on Figure \ref{pic for the Z problem} and given by\footnote{Notice that we can equivalently write \( s_1=-\hat \delta(1)^{-2}(1+e^{-\pi\ic\nu})e^{-2\ic t} \) and \( s_2 = \hat \delta(1)^2 (1-e^{-\pi\ic\nu})e^{-2\ic t} \).}
    \begin{align} \label{defn of s1 and s2}
    \begin{aligned}
        &H_0^Z(\zeta) = \begin{pmatrix}
        1 & 0\\
        - s_2 e^{2 \pi \ic \nu} e^{-\zeta^2/2} & 1
        \end{pmatrix}, \quad
        H_1^Z(\zeta) = \begin{pmatrix}
        1 & s_1 e^{\zeta^2/2}\\
        0 & 1
        \end{pmatrix}, \\
        &H_2^Z(\zeta) = \begin{pmatrix}
        1 & 0\\
        s_2 e^{-\zeta^2/2} & 1
        \end{pmatrix}, \quad
        H_3^Z(\zeta) = \begin{pmatrix}
        1 & - s_1 e^{2\pi \ic \nu} e^{\zeta^2/2}\\
        0 & 1
        \end{pmatrix}, \\
        &D^Z = e^{2 \pi \ic \nu \sigma_3}, \quad s_1 = - \frac{1}{\hat \delta(1)^2} \frac{h(1) + 1}{h(1)} e^{-2\ic t}, \quad s_2 = \hat \delta(1)^2\frac{h(1) - 1}{h(1)} e^{2 \ic t};
    \end{aligned}
    \end{align}
    \item it holds that \( Z(\zeta) = (I + \O(1/\zeta)) \zeta^{-\nu \sigma_3} \), \( \zeta \to \infty \), where \( \zeta^\nu>0 \) when \( \zeta>0 \) and \( \arg(\zeta)\in(-\pi/4,7\pi/4) \).
\end{enumerate}
\end{RHP}

This is the same setup as described in \cite[RHP 7]{IMY1} (and many other papers). RHP~\ref{parabolic cylinder rhp} can be solved explicitly, see \cite[page 21]{IMY1}. Indeed, let \( \arg(\zeta)\in(-\frac\pi 4,\frac {7\pi}4 ) \) and set
\begin{align}
\Omega_0 &= \left\{ \zeta \in \C \,|\, \arg(\zeta)\in\left(-\tfrac\pi4,0\right) \right\}, & \; \Omega_1 = \left\{ \zeta \in \C \,|\,\;\; \arg(\zeta)\in\left(0,\tfrac\pi2\right) \right\},\\ 
\Omega_2 &= \left\{ \zeta \in \C \,|\, \arg(\zeta)\in\left(\tfrac\pi2, \pi\right) \right\}, & \; \Omega_3 = \left\{ \zeta \in \C \,|\, \arg(\zeta)\in\left(\pi,\tfrac{3\pi}{2}\right) \right\},\\
\Omega_4 &= \left\{ \zeta \in \C \,|\, \arg(\zeta)\in\left(\tfrac{3\pi}{2}, \tfrac{7\pi}4\right) \right\}.
\end{align}
Then, the solution of RHP~\ref{parabolic cylinder rhp} is given by
\begin{equation} 
\label{defn of Z}
Z(\zeta)e^{\frac14\zeta^2\sigma_3} 
=\begin{cases}
    \begin{pmatrix} D_{-\nu}\left(\zeta e^{\pi \ic/2} \right )e^{\pi \ic\nu /2} & -\hat\alpha D_{\nu-1}(\zeta) \\ \ic \hat\beta D_{-\nu-1}\left(\zeta e^{\pi \ic/2}\right )e^{\pi \ic \nu /2} & D_{\nu}(\zeta) \end{pmatrix}, & \zeta \in \Omega_0, \smallskip \\
    \begin{pmatrix} D_{-\nu}\left(\zeta e^{-\pi \ic / 2} \right )e^{-\pi \ic \nu /2} & -\hat\alpha D_{\nu-1}(\zeta) \\ -\ic \hat\beta D_{-\nu-1}\left(\zeta e^{-\pi \ic / 2}\right )e^{-\pi \ic \nu /2} & D_{\nu}(\zeta) \end{pmatrix}, & \zeta \in \Omega_1, \smallskip \\
    \begin{pmatrix} D_{-\nu}\left(\zeta e^{-\pi \ic/2} \right )e^{-\pi \ic \nu /2} & \hat\alpha D_{\nu-1}\left(\zeta e^{-\pi \ic} \right)e^{\pi \ic \nu} \\ -\ic\hat\beta D_{-\nu-1}\left(\zeta e^{-\pi \ic/2}\right )e^{-\pi \ic\nu /2} & D_{\nu}\left(\zeta e^{-\pi \ic} \right)e^{\pi \ic \nu} \end{pmatrix}, & \zeta \in \Omega_2, \smallskip \\
    \begin{pmatrix} D_{-\nu}\left(\zeta e^{-3\pi \ic/2} \right )e^{-3\pi \ic\nu/2} & \hat\alpha D_{\nu-1}\left(\zeta e^{-\pi \ic} \right)e^{\pi \ic \nu} \\ \ic \hat\beta D_{-\nu-1}\left(\zeta e^{-3\pi \ic/2}\right )e^{-3\pi \ic \nu /2} & D_{\nu}\left(\zeta e^{-\pi \ic} \right)e^{\pi \ic \nu} \end{pmatrix}, & \zeta \in \Omega_3, \smallskip \\
    \begin{pmatrix} D_{-\nu}\left(\zeta e^{-3\pi \ic/2} \right )e^{-3\pi \ic \nu /2} & -\hat\alpha D_{\nu-1}\left(\zeta e^{-2\pi \ic} \right)e^{2\pi \ic \nu} \\ \ic \hat\beta D_{-\nu-1}\left(\zeta e^{-3\pi \ic/2}\right )e^{-3\pi \ic \nu /2} & D_{\nu}\left(\zeta e^{-2\pi \ic} \right)e^{2\pi \ic \nu} \end{pmatrix}, & \zeta \in \Omega_4.
\end{cases}
\end{equation}
Thus, one can refine the asymptotics of $Z(\zeta)$ as $\zeta \to \infty$ as
\begin{equation}
\label{matrix Z}
Z(\zeta) = \left( I + \frac{1}{\zeta} \begin{pmatrix} 0 & -\hat\alpha \\ \hat\beta & 0 \end{pmatrix} + \frac{1}{\zeta^2} \begin{pmatrix} \frac12\nu(1+\nu) & 0 \\ 0 & \frac12\nu(1-\nu) \end{pmatrix} + \O\left(\frac{1}{\zeta^3}\right) \right) \zeta^{-\nu\sigma_3},
\end{equation}
where \( \zeta^\nu>0 \) when \( \zeta>0 \) and \( \arg(\zeta)\in(-\pi/4,7\pi/4) \),
\begin{align} 
\label{alpha and beta}
\hat\alpha = -e^{-2\pi \ic \nu}\frac{\ic}{s_2}\frac{\sqrt{2\pi}}{\Gamma(\nu)} \quad \text{and} \quad \hat\beta = -\frac{\nu}{\hat\alpha}.
\end{align}
Since \( \nu \) is purely imaginary, the entries of \( Z(\zeta) \) are entire functions, and \eqref{matrix Z} holds uniformly for all \( |\zeta| \) large, we have that
\begin{align}
\label{Z_- is bounded}
|[Z_-(\zeta)]_{ij}|\leq M, \quad \zeta\in\R\cup\ic\R,
\end{align}
for some constant \( M \) and all \( i,j\in\{1,2\} \) (this can be independently verified by using \cite[Equation (12.2.5), (12.5.3), and (12.5.5)]{NIST:DLMF}).


\subsection{Approximate Local Parametrices at $w = \pm1$}

As we indicated before, we use the following conformal map to set up a correspondence between $w$ and $\zeta$ planes:
\begin{align} 
\label{zeta <-> w}
    \zeta (w) &= e^{\pi \ic /2} \sqrt{\varphi(w) - 2\ic} = e^{3\pi \ic/4} \, \frac{w-1}{\sqrt{w}}, \quad w\in\D_\epsilon(1),
\end{align}
where \( \sqrt w \) is the principal branch and the other root is defined so that the second equality holds. Notice that the interval \( (1-\epsilon,1) \) is mapped by \( \zeta(w) \) into the ray \( \arg(\zeta)=-\pi/4 \). Since we had quite a bit of freedom in choosing the arcs \( \Gamma^\pm_{ext} \) and \( \Gamma^\pm_{int} \), we fix them around \( 1 \) so that \( \zeta \) maps the arcs \( \Gamma_i \), see Figure~\ref{Y tilde jump matrices}, into the coordinate axes. Then, the map \( \sqrt{2t}\zeta(w) \) maps the jump contour on Figure~\ref{contour and matrix for Y hat} into \( \Gamma_Z \) from Figure~\ref{pic for the Z problem} for any \( t>0 \) so that the matrix \( \hat H_i \) corresponds to the matrix \( H_i^Z \) for any \( i\in\{0,1,2,3\} \). That is, the matrix
\[
Z\big( \sqrt{2t} \zeta(w) \big)
\]
has the same jumps as \( \hat Y(w,t) \) except for \( \hat \delta(w) \) and \( h(w) \) being replaced by \( \hat \delta(1) \) and \( h(1) \), respectively. Now, by undoing the transformation \eqref{defn of Y hat near 1} and \eqref{defn of Y tilde near 1} we arrive at an approximate local parametrix
\begin{align} 
\label{defn of P (1)}
P^{(1)}(w, t) = V^{(1)}(w, t) Z\big( \sqrt{2t} \zeta(w) \big) (w - 1)^{\nu \sigma_3} B, \quad w \in \D_\epsilon(1),
\end{align}
where \( V^{(1)}(w, t) \) could be any matrix function holomorphic in \( \D_\epsilon(1) \). We choose this holomorphic prefactor to ensure good matching with the global parametrix and set
\begin{align}
\begin{aligned}
\label{defn of V^(1)}
V^{(1)}(w, t) & = A(w) \big(\sqrt{2t} \zeta(w) \big)^{\nu \sigma_3} (w-1)^{-\nu \sigma_3} \\  &= A(w) \big(\sqrt{2t} e^{3\pi\ic/4} w^{-1/2} \big)^{\nu \sigma_3},
\end{aligned}
\end{align}
where \( A(w) \) was defined in \eqref{defn of P gl} and \( w^{\nu/2} \) is the principal branch.

Due to \eqref{sigma1 symmetry of Y}, we define the local parametrix \( P^{(-1)}(w) \) near $w = -1$ by setting
\begin{align} 
\label{defn of P-1}
    P^{(-1)}(w, t) = \sigma_1 P^{(1)}(-w, t) \sigma_1 \begin{dcases}
        1, & w\in\mathrm{ext}(\Gamma_{gl}),\\
        -1, & w\in\mathrm{int}(\Gamma_{gl}).
    \end{dcases}
\end{align}
If we denote the jump matrices for \( P^{(\pm1)}(w) \) by \( G_{P^{(\pm1)}} (w, t) \), then it holds that
\begin{align}
\label{jump P+-1 symmetry}
G_{P^{(-1)}} (w, t) = \sigma_1G_{P^{(1)}} (-w, t) \sigma_1.
\end{align}


\subsection{Comparison between RHP~\ref{rhp Y} and Approximate Local Parametrices}

For future use we now compare \( Y(w,t) \) to the local paramatrices \( P^{(\pm1)}(w)\) in \( \D_\epsilon(\pm1) \). These matrices have matching jumps on \( \Gamma_{gl} \cap \D_\epsilon(1) \) while
\begin{align}
\label{G_Y explicit}
G_Y(w, t) = 
\begin{cases}
\begin{pmatrix}
    1 & \hat \delta(w)^2 (w - 1)^{2\nu} e^{2\pi \ic \nu} \frac{h(w) - 1}{h(w)} e^{t\varphi(w)}\\
    0 & 1
\end{pmatrix}, & w \in \Gamma_0, \smallskip \\
\begin{pmatrix}
    1 & \frac{-1}{\hat \delta(w)^2} (w - 1)^{-2\nu} \frac{h(w) + 1}{h(w)} e^{-t\varphi(w)}\\
    0 & 1
\end{pmatrix}, & w \in \Gamma_1, \smallskip \\
\begin{pmatrix}
    1 & 0\\
    \hat \delta(w)^2 (w - 1)^{2\nu} \frac{h(w) - 1}{h(w)} e^{t\varphi(w)} & 1
\end{pmatrix}, & w \in \Gamma_2, \smallskip \\
\begin{pmatrix}
    1 & 0\\
    \frac{-1}{\hat \delta(w)^2} (w - 1)^{-2\nu} e^{2\pi \ic \nu} \frac{h(w) + 1}{h(w)} e^{-t\varphi(w)} & 1
\end{pmatrix}, & w \in \Gamma_3,
\end{cases}
\end{align}
by \eqref{defn of H0 tilde - H3 tilde}, \eqref{delta of w near w=1}, and the definition of \( B \) in \eqref{defn of P gl}. Since the jump matrix for \( P^{(1)}(w) \) is obtained from \( G_Y(w, t) \) by replacing \( \hat \delta(w), h(w) \) with \( \hat \delta(1), h(1) \) we have that
\begin{align} \label{GYGP1^-1}
&\qquad G_{Y} (w, t) G_{P^{(1)}} (w, t)^{-1}\\
&= 
\begin{cases}
\begin{pmatrix}
    1 & (w - 1)^{2\nu} e^{2\pi \ic \nu} \left( \hat \delta(w)^2 \frac{h(w) - 1}{h(w)} - \hat \delta(1)^2 \frac{h(1) - 1}{h(1)} \right) e^{t\varphi(w)}\\
    0 & 1
\end{pmatrix}, & w \in \Gamma_0, \smallskip \\
\begin{pmatrix}
    1 & (w - 1)^{-2\nu} \left( \frac{1}{\hat \delta(1)^2} \frac{h(1) + 1}{h(1)} - \frac{1}{\hat \delta(w)^2} \frac{h(w) + 1}{h(w)} \right) e^{-t\varphi(w)}\\
    0 & 1
\end{pmatrix}, & w \in \Gamma_1, \smallskip \\
\begin{pmatrix}
    1 & 0\\
    (w - 1)^{2\nu} \left( \hat \delta(w)^2 \frac{h(w) - 1}{h(w)} - \hat \delta(1)^2 \frac{h(1) - 1}{h(1)} \right) e^{t\varphi(w)} & 1
\end{pmatrix}, & w \in \Gamma_2, \smallskip \\
\begin{pmatrix}
    1 & 0\\
    (w - 1)^{-2\nu} e^{2\pi \ic \nu} \left( \frac{1}{\hat \delta(1)^2} \frac{h(1) + 1}{h(1)} - \frac{1}{\hat \delta(w)^2} \frac{h(w) + 1}{h(w)} \right) e^{-t\varphi(w)} & 1
\end{pmatrix}, & w \in \Gamma_3.
\end{cases}
\end{align}

In another connection, we get on \( \partial \D_\epsilon(1) \) that
\begin{multline} 
\label{matching cond 1 ver 0}
    A(w)^{-1} P^{(1)}(w, t) P^{(gl)}(w)^{-1} A(w)= I + \frac{1}{\sqrt{2t} \zeta(w)} \begin{pmatrix}
        0 & - \tilde \alpha(w)\\
        \tilde \beta(w) & 0
    \end{pmatrix} \\
     + \frac{1}{2t\zeta(w)^2} \begin{pmatrix}
        \frac{1}{2}\nu(1+\nu) & 0\\
        0 & \frac{1}{2}\nu(1 - \nu)
    \end{pmatrix} + \O\big(t^{-3/2}\big)
\end{multline}
by \eqref{P global}, \eqref{defn of P (1)}, \eqref{defn of V^(1)}, and \eqref{matrix Z} where
\begin{align} 
\label{alpha tilde and beta tilde}
    \tilde \alpha (w) := \big(2t e^{3\pi\ic/2}/w \big)^\nu \hat\alpha, \quad \tilde \beta (w) := \big(2t e^{3\pi\ic/2}/w \big)^{-\nu} \hat\beta.
\end{align}
In particular, we get from the definition of \( A(w) \) in \eqref{defn of P gl} that
\begin{multline} 
\label{matching cond 1}
\left[P^{(1)}(w, t) P^{(gl)}(w)^{-1}\right]_{11} = 1 + \frac{1}{\sqrt{2t} \zeta(w)} \left( \frac{w\tilde \beta (w)}{(w+\ic)^2}  - \frac{w\tilde \alpha (w)}{(w - \ic)^2}  \right)\\
+\frac{1}{2t \zeta(w)^2} \frac\nu2 \frac{w^2(1+\nu)+(1-\nu)}{w^2 + 1} + \O\big(t^{-3/2}\big),   
\end{multline}
and
\begin{multline} 
\label{matching cond 2}
\left[P^{(1)}(w, t) P^{(gl)}(w)^{-1}\right]_{22} = 1 - \frac{1}{\sqrt{2t} \zeta(w)} \left( \frac{w\tilde \beta (w)}{(w+\ic)^2}  - \frac{w\tilde \alpha (w)}{(w - \ic)^2}  \right)\\
+\frac1{2t \zeta(w)^2} \frac\nu2 \frac{w^2(1-\nu)+(1+\nu)}{w^2 + 1} + \O\big(t^{-3/2}\big).
\end{multline}
Clearly, analogous results in \( \D_\epsilon(-1) \) can be deduced from the symmetry relations \eqref{sigma1 symmetry of Y} and \eqref{sigma1 symmetry of Pgl} as well as the definition \eqref{defn of P-1}.


\section{Small Norm Problem}

\begin{figure}[htbp]
    \centering
    \includegraphics[width=5cm]{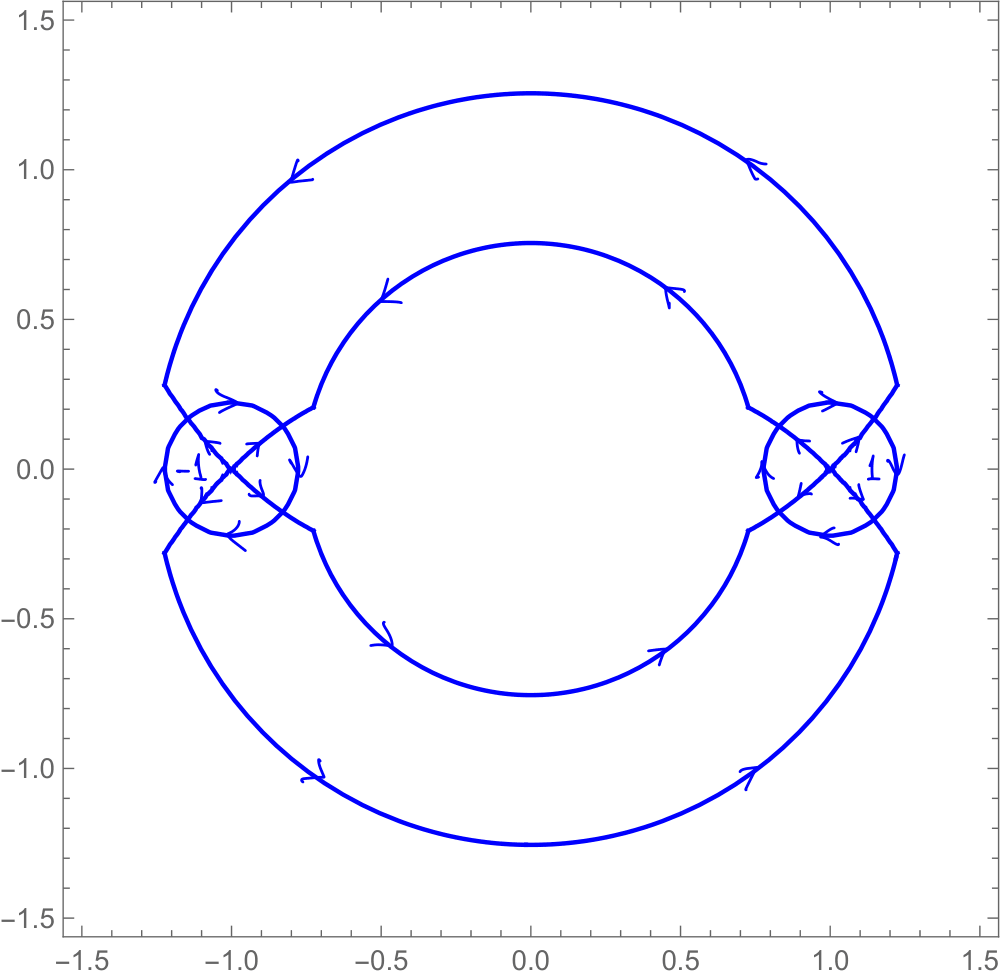}
    \caption{The contour $\Gamma_R$ for RHP~\ref{rhp R}.}
    \label{pic for the R problem}
\end{figure}

We look for the solution of RHP~\ref{rhp Y} in the form
\begin{align} \label{defn of R}
    Y(w) = R(w) \begin{cases}
        P^{(gl)}(w), \quad w \in \C \setminus (\Gamma_{gl} \cup \D_\epsilon(1) \cup \D_\epsilon(-1)), \\
        P^{(\pm 1)}(w), \quad w \in \D_\epsilon(\pm 1).
    \end{cases}
\end{align}
Then, the error function $R(w)$ must solve the following Riemann-Hilbert problem.
\begin{RHP}\label{rhp R}
Find a $2 \times 2$ matrix function $R(w, t)$ such that
\begin{enumerate}
    \item $R(w)$ is analytic in $\C \setminus \Gamma_R$, where \( \Gamma_R = \Gamma_R^{out} \cup \Gamma_R^{bnd} \cup \Gamma_R^{in} \), see Figure \ref{pic for the R problem}, with
    \begin{align}
        \Gamma_R^{out} &:= \Gamma \setminus \left( \Gamma_{gl} \cup \D_\epsilon(1) \cup \D_\epsilon(-1) \right),\\
        \Gamma_R^{bnd} & := \partial \D_\epsilon(1) \cup \partial \D_\epsilon(-1), \\
        \Gamma_R^{in} &:= \left( (\Gamma\setminus\Gamma_{gl}) \cap \D_\epsilon(1) \right) \cup \left( (\Gamma\setminus\Gamma_{gl}) \cap \D_\epsilon(-1) \right); 
    \end{align}
    \item one-sided traces $R_\pm (w)$ exist a.e. on \( \Gamma_R \), are bounded there, and satisfy
    \begin{align*}
        R_+(w, t) = R_-(w, t) G_{R}(w, t), \quad w \in \Gamma_R,
    \end{align*}
    where
    \begin{equation} 
    \label{defn of GR}
        \quad G_R(w, t) = \begin{dcases}
            P^{(gl)}(w) G_{Y}(w, t) P^{(gl)}(w)^{-1}, & w \in \Gamma_R^{out},\\
            P_-^{(\pm 1)}(w, t) G_{Y}(w, t) G_{P^{(\pm 1)}}(w, t)^{-1} P_-^{(\pm 1)}(w, t)^{-1}, & w \in \Gamma_R^{in},\\
            P^{(\pm 1)}(w, t) P^{(gl)}(w)^{-1}, & w \in \Gamma_R^{bnd};
        \end{dcases}
    \end{equation}
    \item it holds that $R(w) = I + w^{-1} R_1(t) + \O \left(w^{-2} \right)$ as $w \to \infty$.
\end{enumerate}
\end{RHP}
In what follows from \eqref{GY symmetry}, \eqref{sigma1 symmetry of Pgl}, \eqref{defn of P-1}, and \eqref{jump P+-1 symmetry} that
\begin{align}
\label{E symmetry}
E(w)= \sigma_1 E(-w) \sigma_1, \quad E(w):= G_R(w)-I.
\end{align}

\subsection{Integral Representation}

Since \( \hat\delta(w),h(w)\), and \( (w-1)^\nu \) are fixed analytic functions, and so are the entries of \( P^{(gl)}(w) \), it follows from \eqref{defn of L1 P1 L2 U1 P2 U2}, \eqref{sign chart}, and \eqref{defn of G Y} that
\begin{align} 
\label{G R - I exp small outside}
    \| E \|_{L^{\infty} (\Gamma_R^{out})} = \O(e^{-c_{\epsilon} t}),
\end{align}
where $c_\epsilon$ is some constant dependent on the radius $\epsilon$ of the chosen disks around $w = \pm 1$. On $\Gamma_R^{bnd}$, we have from \eqref{defn of P-1} and \eqref{matching cond 1 ver 0} that
\begin{align} 
\label{G R - I small boundary}
    \| E \|_{L^{\infty}(\Gamma_R^{bnd})} = \O \big( t^{-1/2} \big).
\end{align}
Next, due to the choice of the arcs \( \Gamma_{int}^\pm \) and \( \Gamma_{ext}^\pm \) made right after \eqref{zeta <-> w} and since \( \hat\delta(w) \) and \( h(w) \) are differentiable functions on those arcs, see \eqref{delta'(w)}, we see that off-diagonal entries of the matrices in \eqref{GYGP1^-1} are of the form
\[
\{\text{bounded function}\}\times \zeta e^{-t\zeta^2} 
\]
for \( \zeta>0 \). Hence, we have that
\[
E(w) = \{\text{bounded function}\}\times \zeta(w) e^{-t\zeta^2(w)} P_-^{(1)}(w) \begin{pmatrix} 0 & 1 \\ 0 & 0 \end{pmatrix} P_-^{(1)}(w)^{-1}
\]
for \( w\in\Gamma_0 \) and \( w\in\Gamma_1 \). Clearly, similar expressions can be written for \( w\in\Gamma_2 \) and \( w\in\Gamma_3 \). Thus, we can see from \eqref{Z_- is bounded}, \eqref{defn of P (1)}, \eqref{defn of V^(1)}, and \eqref{defn of P-1} that
\[
E(w) = \zeta(w) e^{-t\zeta^2(w)}\O(1), \quad w\in\Gamma_R^{in},
\]
where the entries of \( \O(1) \) are bounded functions. Since \( 1/\zeta^\prime(w)\) is a bounded function in \( \D_\epsilon(1) \) as well,  a trivial calculus computation shows that 
\begin{align}
\label{G R - I small inside}
\|E\|_{L^p(\Gamma_R^{in})} = \O\big(t^{-1/2-1/(2p)}\big)
\end{align}
for any \( p\in[1,\infty] \), where the estimate in \( \D_\epsilon(-1)\) follows from symmetry \eqref{E symmetry}. Altogether, we get that
\begin{align} 
\label{G R - I small}
\| E \|_{L^{\infty}(\Gamma_R)} = \O \big( t^{-1/2} \big) \qasq t\to\infty.
\end{align}
Hence, by the small norm theorem, see for example \cite[Theorem~8.1]{FIKN}, there exists a unique solution of RHP~\ref{rhp R} for all $t$ large and
\begin{align}\label{singular int eq}
    R(w) = I + \frac{1}{2 \pi \ic} \int_{\Gamma_R} \frac{\rho(w')E(w')}{w' - w} dw', \quad w \notin \Gamma_{R}, 
\end{align}
where $\rho(w') = R_-(w')$ for $w' \in \Gamma_R$. Let $(\mathcal{C}_EF)(w) := (\mathcal{C}(FE))_-(w)$, \( w\in\Gamma_R \), where $\mathcal C$ is the Cauchy operator on \( \Gamma_R \). Then, 
\[
\rho(w) = I + (\mathcal{C}_E\rho)(w), \quad w\in\Gamma_R,
\]
which is a singular integral equation obtained by taking the traces of both sides of \eqref{singular int eq} on the left-hand side of \(\Gamma_R\). By \eqref{G R - I small}, \( \|\mathcal C_E\|=\O(t^{-1/2})\) and therefore it holds that
\begin{equation}
\label{von Neumann}
\rho = (\mathcal{I} - \mathcal{C}_E)^{-1} (I) =  \sum_{m=0}^\infty \mathcal{C}_E^m I.
\end{equation}
Thus, \eqref{singular int eq} can be written as
\begin{align} \label{S series}
    R(w) &= I + \sum_{m = 0}^\infty \mathcal{C} \left( \left( \mathcal{C}_E^m I \right) E \right) (w) =: \sum_{k=0}^{\infty} \mathcal{S}^{(k)}(w),
\end{align}
where $\mathcal{S}^{(0)}(w) = I$ and $\mathcal{S}^{(k+1)}(w) = \mathcal{C} \big( \mathcal{S}^{(k)}_- E \big)(w)$ for $k \geq 0$. In particular,
\begin{align}
\label{R1 11 component eq 1}
    [R_1(t)]_{11} =  \lim_{w \to \infty} w\big[ R(w) - I \big]_{11} = \sum_{k=1}^{\infty} \lim_{w \to \infty} w  \big[\mathcal{S}^{(k)}(w) \big]_{11},
\end{align}
where we used \eqref{S series} for the last equality. Our goal is to asymptotically evaluate the above quantity as $t \to \infty$.


\subsection{Asymptotic Analysis of \( \mathcal S^{(1)} \)} 
\label{AAIII}

By its very definition, we have that
\begin{align}
\label{limit S1 prelim}
\lim_{w \to \infty} w \big[ \mathcal{S}^{(1)}(w) \big]_{11} = -\frac{1}{2\pi \ic} \int_{\Gamma_R}[E(w')]_{11} dw =  I_{out} + I_{bnd} + I_{in},
\end{align}
where the summands on the right-hand side above represent integrals over \( \Gamma_R^{out} \), \( \Gamma_R^{bnd} \), and \( \Gamma_R^{in} \), respectively. Due to \eqref{G R - I exp small outside}, we have that
\begin{equation}
\label{I-out}
I_{out} = -\frac{1}{2\pi \ic} \int_{\Gamma_R^{out}} [E(w')]_{11} dw = \O(e^{-c_\epsilon t}).
\end{equation}

Now, it follows from \eqref{E symmetry} that
\begin{align}
I_{bnd}  = -\frac{1}{2\pi \ic} \int_{\partial \D_\epsilon(1)} \big(\left[E(w)\right]_{11} - \left[E(w)\right]_{22}\big) dw.
\end{align}
According to \eqref{matching cond 1} and \eqref{matching cond 2} the integrand above is equal to
\[
\frac{\sqrt2}{\sqrt{t} \zeta(w)} \left( \frac{w\tilde \beta (w)}{(w+\ic)^2} - \frac{w\tilde \alpha (w)}{(w - \ic)^2}  \right) + \frac{\nu^2}{2t \zeta(w)^2} \frac{w^2-1}{w^2 + 1} + \O\big(t^{-3/2}\big).
\]
Recall that both circles \( \partial \D_\epsilon(\pm1) \) are oriented clockwise. Recall also \eqref{zeta <-> w} and \eqref{alpha tilde and beta tilde}. Thus, by the Cauchy integral formula, one has that
\begin{align}
I_{bnd} &= e^{3\pi\ic/4} \frac{\tilde \alpha(1)+\tilde\beta(1)}{\sqrt{2t}} + \frac{\ic \nu^2}{2t} + \O\big(t^{-3/2}\big) \\
\label{I-bnd}
& = d_1 e^{-2\ic t + \nu \ln t} t^{-1/2} + d_2 e^{2\ic t - \nu \ln t} t^{-1/2} + \frac{\ic \nu^2}{2t} + \O\big(t^{-3/2}\big),
\end{align}
where, in view of \eqref{alpha tilde and beta tilde}, \eqref{alpha and beta}, and \eqref{defn of s1 and s2}, 
\begin{align}
\label{d1 and d2}
d_1 = e^{\pi \ic(1- 2\nu)/4} \frac{2^{\nu}}{\hat{\delta}(1)^2} \frac{h(1)}{h(1) - 1} \frac{\sqrt{\pi}}{\Gamma(\nu)} \qandq  d_2 = \frac{\ic\nu}{2d_1}. 
\end{align}

Next, we deduce from \eqref{E symmetry} that
\begin{align}
\label{target integral in}
I_{in} = \frac{1}{2\pi \ic} \int_{\Gamma_R^{in}\cap \D_\epsilon(1)} \big( [E(w)]_{22} - [E(w)]_{11} \big) dw.
\end{align}
To compute this integral, one needs an explicit expression of 
\[
G_R(w,t) = P_{-}^{(1)}(w,t) G_Y(w,t) G_{P^{(1)}}^{-1}(w,t) P_{-}^{(1)}(w,t)^{-1}
\]
for $w \in \Gamma_R^{in}\cap\D_\epsilon(1)$. We need to introduce more notation. Recall \eqref{defn of Z}. We shall write
\begin{align} \label{Z entrywise defn}
Z_- \left(\sqrt{2t} \zeta(w)\right) = \begin{pmatrix}
        f_1^{(i)} \left(\sqrt{2t} \zeta(w) \right) & \hat \alpha f_2^{(i)} \left(\sqrt{2t} \zeta(w) \right) \smallskip \\
        \hat \beta f_3^{(i)} \left(\sqrt{2t} \zeta(w) \right) & f_4^{(i)} \left(\sqrt{2t} \zeta(w) \right)
    \end{pmatrix}, \quad w\in \Gamma_i,
\end{align}
for \( i\in\{0,1,2,3\} \). Further, we write
\[
P^{(1)}_-(w,t) =: \begin{pmatrix}
        a^{(i)}(w,t) & b^{(i)}(w,t)\\
        c^{(i)}(w,t) & d^{(i)}(w,t)
    \end{pmatrix}, \quad w\in\Gamma_i,
\]
again, for \( i\in\{0,1,2,3\} \). Then, according to \eqref{defn of P gl}, \eqref{defn of P (1)}, and \eqref{defn of V^(1)} we have that
\[
a^{(i)}(w,t) = \begin{cases}
\displaystyle \frac{\hat\alpha w}{w-\ic} \frac{(*)^\nu}{(w-1)^{2\nu}} f_2^{(i)}(*) + \frac1{w+\ic}  (*)^{-\nu}  f_4^{(i)}(*), & i\in\{0,3\}, \smallskip \\
\displaystyle \frac{w}{w-\ic} (*)^\nu f_1^{(i)}(*) + \frac{\hat\beta}{w+\ic}  \frac{(w-1)^{2\nu}}{(*)^\nu}  f_3^{(i)}(*), & i\in\{1,2\},
\end{cases}
\]
where, for brevity, \( * = \sqrt{2t}\zeta(w) \),
\[
b^{(i)}(w,t) = \begin{cases}
\displaystyle \frac{-w}{w-\ic} (*)^\nu f_1^{(i)}(*) - \frac{\hat\beta}{w+\ic}  \frac{(w-1)^{2\nu}}{(*)^\nu}  f_3^{(i)}(*), & i\in\{0,3\}, \smallskip \\
\displaystyle \frac{\hat\alpha w}{w-\ic} \frac{(*)^\nu}{(w-1)^{2\nu}} f_2^{(i)}(*) + \frac1{w+\ic}  (*)^{-\nu}  f_4^{(i)}(*), & i\in\{1,2\},
\end{cases}
\]
and
\[
c^{(i)}(w,t) = \begin{cases}
\displaystyle \frac{-\hat\alpha}{w-\ic} \frac{(*)^\nu}{(w-1)^{2\nu}} f_2^{(i)}(*) + \frac w{w+\ic}  (*)^{-\nu}  f_4^{(i)}(*), & i\in\{0,3\}, \smallskip \\
\displaystyle \frac{-1}{w-\ic} (*)^\nu f_1^{(i)}(*) + \frac{\hat\beta w}{w+\ic}  \frac{(w-1)^{2\nu}}{(*)^\nu}  f_3^{(i)}(*), & i\in\{1,2\},
\end{cases}
\]
while
\[
d^{(i)}(w,t) = \begin{cases}
\displaystyle \frac{1}{w-\ic} (*)^\nu f_1^{(i)}(*) - \frac{\hat\beta w}{w+\ic}  \frac{(w-1)^{2\nu}}{(*)^\nu}  f_3^{(i)}(*), & i\in\{0,3\}, \smallskip \\
\displaystyle \frac{-\hat\alpha}{w-\ic} \frac{(*)^\nu}{(w-1)^{2\nu}} f_2^{(i)}(*) + \frac w{w+\ic}  (*)^{-\nu}  f_4^{(i)}(*), & i\in\{1,2\}.
\end{cases}
\]
Denote the off-diagonal entry of \( G_Y(w,t) G_{P^{(1)}}^{-1}(w,t) \) on \( \Gamma_i \) by \( g^{(i)}(w,t) \), see \eqref{GYGP1^-1}. Then,
\begin{align*}
    g^{(0)}(w,t) &= e^{2\pi\ic\nu} g^{(2)}(w,t) = (w-1)^{2\nu}e^{2\pi \ic \nu}\Big(\hat{\delta}(w)^{2}\tfrac{h(w)-1}{h(w)}-\hat{\delta}(1)^{2}\tfrac{h(1)-1}{h(1)}\Big)e^{t\varphi(w)},\\
    g^{(3)}(w,t) &= e^{2\pi\ic\nu} g^{(1)}(w,t)= (w-1)^{-2\nu}e^{2\pi \ic \nu}\Big( \tfrac1{\hat\delta(1)^2}\tfrac{h(1)+1}{h(1)} - \tfrac1{\hat\delta(w)^2}\tfrac{h(w)+1}{h(w)} \Big)e^{-t\varphi(w)}.
\end{align*}
Recall that all our matrices have determinant 1. Thus, we have
\[
[E(w)]_{22} - [E(w)]_{11} = 2g^{(i)}(w,t)\begin{cases}
    a^{(i)}(w,t) c^{(i)}(w,t), & i\in\{0,1\}, \\
    -b^{(i)}(w,t) d^{(i)}(w,t), & i\in\{2,3\}.
 \end{cases}
\]
If we set
\[
\ell^{(i)}(w,t) := \begin{cases}
a^{(0)}(w,t) c^{(0)}(w,t), & i=0, \\
a^{(1)}(w,t) c^{(1)}(w,t), & i=1, \\
b^{(2)}(w,t) d^{(2)}(w,t), & i=2, \\
b^{(3)}(w,t) d^{(3)}(w,t), & i=3,
\end{cases}
\]
then it holds that
\begin{equation}
\label{defn of k}
\ell^{(i)}(w,t) = -\hat\alpha^2\frac{ w}{(w-\ic)^2}\frac{(*)^{2\nu}f_2^{(i)}(*)^2}{(w-1)^{4\nu}} + \hat\alpha \frac{w^2-1}{w^2+1} \frac{f_2^{(i)}(*)f_4^{(i)}(*)}{(w-1)^{2\nu}} + \frac w{(w+\ic)^2} \frac{f_4^{(i)}(*)^2}{(*)^{2\nu}}
\end{equation}
when \( i\in\{0,2\} \), where again \( * = \sqrt{2t}\zeta(w) \), and
\begin{multline}
\ell^{(i)}(w,t) = -\frac{w}{(w-\ic)^2}(*)^{2\nu}f_1^{(i)}(*)^2 + \hat\beta \frac{w^2-1}{w^2+1} (w-1)^{2\nu} f_1^{(i)}(*)f_3^{(i)}(*) \\ + \hat\beta^2 \frac w{(w+\ic)^2} \frac{(w-1)^{4\nu}}{(*)^{2\nu}}f_3^{(i)}(*)^2
\end{multline}
when \( i\in\{1,3\} \). We can now rewrite \eqref{target integral in} as
\begin{multline} 
\label{target integral ver 2}
I_{in} = \frac{1}{\pi \ic} \int_{\Gamma_0} g^{(0)}(w,t)\ell^{(0)}(w,t) dw - \frac{1}{\pi \ic} \int_{\Gamma_2} g^{(2)}(w,t)\ell^{(2)}(w,t) dw\\
+ \frac{1}{\pi \ic}\int_{\Gamma_1} g^{(1)}(w,t)\ell^{(1)}(w,t) dw  - \frac{1}{\pi \ic}\int_{\Gamma_3} g^{(3)}(w,t)\ell^{(3)}(w,t) dw.
\end{multline}

We continue by analyzing the first integral in \eqref{target integral ver 2}. The product of \( g^{(0)}(w,t) \) and the middle summand in \eqref{defn of k} can be written as
\[
2 \ic \hat\alpha e^{2\pi\ic\nu+2\ic t} f_2^{(0)}\big(\sqrt{2t}\zeta(w)\big) f_4^{(0)}\big(\sqrt{2t}\zeta(w)\big) \zeta(w)\zeta^\prime(w)F(w)e^{-t\zeta^2(w)},
\]
where
\[
F(w) = \frac{w^2}{w^2+1} \Big(\hat{\delta}(w)^{2}\tfrac{h(w)-1}{h(w)}-\hat{\delta}(1)^{2}\tfrac{h(1)-1}{h(1)}\Big).
\]
Observe that \( F(w) \) is a differentiable function on \( \Gamma_0 \) that vanishes at \( w=1 \). Recall that \( \zeta(w) \) maps \( \Gamma_0 \) into the positive reals. Then, the integral of interest can be written as
\begin{align}
I_{in}^{(0,2)} & := \frac{2\hat\alpha}\pi e^{2\pi\ic\nu+2\ic t} \int_0^{\zeta_0}f_2^{(0)}\big(\sqrt{2t}\zeta\big) f_4^{(0)}\big(\sqrt{2t}\zeta\big) \zeta F(w(\zeta))e^{-t\zeta^2}d\zeta  \\
& = \frac{\hat\alpha}{\pi t} e^{2\pi\ic\nu+2\ic t}\int_0^{\sqrt{2t}\zeta_0} f_2^{(0)}(s)f_4^{(0)}(s)  F\big(w(s/\sqrt{2t})\big) s e^{-s^2/2} ds,
\end{align}
where \( \zeta_0 \) is the value of \( \zeta(w) \) at the point where \( \Gamma_0 \) meets \( \partial \D_\epsilon(1) \). Observe that \( \hat\alpha e^{2\ic t} \) is in fact independent of \( t \), see \eqref{alpha and beta} and \eqref{defn of s1 and s2}. Recall that the functions \( f_2^{(0)}(s)\)  and \( f_4^{(0)}(s) \) are bounded on the positive reals, see \eqref{Z_- is bounded}. Then,
 \begin{align}
 \label{I-in-02}
 \big|I_{in}^{(0,2)}\big| \leq M t^{-3/2} \int_0^{\sqrt{2t}\zeta_0} s^2 e^{-s^2/2} ds = \O\big(t^{-3/2}\big) \qasq t\to\infty
 \end{align}
 for some constant \( M>0 \), where we used estimates \( |F(w)|\leq c|w-1| \) and \( |w(\zeta)|\leq c|\zeta| \) for some constant \( c \) (both functions are continuously differentiable). 

Next, we consider the product of \( g^{(0)}(w,t) \) and the last summand in \eqref{defn of k}. Since this last term does not vanish at \( 1 \), our analysis needs additional steps. This product can be written as
\[
2\frac{e^{\pi\ic\nu/2-3\pi\ic/4+2\ic t}}{(2t)^\nu} f_4^{(0)}\big(\sqrt{2t}\zeta(w)\big)^2 \zeta^\prime(w)G(w)e^{-t\zeta^2(w)}
\]
(notice that we do not have \( \zeta(w) \) in the above formula this time), where
\[
G(w) = \frac{w^{5/2+\nu}}{(w+1)(w+\ic)^2}\Big(\hat{\delta}(w)^{2}\tfrac{h(w)-1}{h(w)}-\hat{\delta}(1)^{2}\tfrac{h(1)-1}{h(1)}\Big).
\]
Because the first factor above is an analytic function in \( \D_\epsilon(1)\), we can, in fact, write
\[
G(w) = \left( \frac1{4\ic} + \zeta(w) G_1(w)\right)\Big(\hat{\delta}(w)^{2}\tfrac{h(w)-1}{h(w)}-\hat{\delta}(1)^{2}\tfrac{h(1)-1}{h(1)}\Big),
\]
where \( G_1(w) \) is an analytic function in \( \D_\epsilon(1)\). Similarly, as we have shown in \eqref{delta'(w)}, the second factor in the formula for \( G(w) \) is a  differentiable function on \( \Gamma_0 \) whose derivative is H\"older continuous with any index \( <1 \). Hence,
\[
G(w) = \zeta(w)\left( \frac1{4\ic} + \zeta(w) G_1(w)\right)\left(e^{-3\pi\ic/4}d +G_2(w)\right),
\]
where \( G_2(w) \) is a continuous function in \( \Gamma_0 \) such that \( |G_2(w)|\leq c_\varepsilon |w-1|^{1-2\varepsilon} \) for any \( \varepsilon>0 \), and \( d \) is the value of the derivative of \( \hat{\delta}(w)^{2}(1-h(w)^{-1}) \) at \( 1 \). Thus,
\begin{align}
I_{in}^{(0,3)} & := 2\frac{e^{\pi\ic\nu/2+3\pi\ic/4+2\ic t}}{\pi (2t)^\nu} \int_0^{\zeta_0} f_4^{(0)}\big(\sqrt{2t}\zeta\big)^2 G(w(\zeta))e^{-t\zeta^2} d\zeta \\
& =  d\frac{e^{\pi\ic\nu/2+2\ic t}}{2\pi\ic (2t)^\nu} \int_0^{\zeta_0} \zeta f_4^{(0)}\big(\sqrt{2t}\zeta\big)^2 e^{-t\zeta^2} d\zeta + \O_\varepsilon \big(t^{-3/2+\varepsilon}\big),
\end{align}
where \( \O_\varepsilon (t^{-3/2+\varepsilon}) \) is obtained as in \eqref{I-in-02}. The integral above can be rewritten as
\begin{align}
\label{integal of D}
\frac1{2t}\int_0^{\sqrt{2t}\zeta_0} sD_\nu^2(s)ds =  \frac1{2t}\int_0^\infty sD_\nu^2(s) ds + o\big(e^{-ct}\big)
\end{align}
for some constant \( c>0 \), see \eqref{defn of Z}. Altogether, we get for some explicit constant \( C_\nu^{(0,3)} \) that
\begin{align}
\label{I-in-03}
I_{in}^{(0,3)} = C_\nu^{(0,3)} e^{2\ic t-\nu\ln t} t^{-1} + \O_\varepsilon (t^{-3/2+\varepsilon}) \qasq t\to\infty.
\end{align}

Now, we consider the product of \( g^{(0)}(w,t) \) and the first summand in \eqref{defn of k}. This product can be written as
\[
-2\hat\alpha^2 e^{7\pi\ic\nu/2-3\pi\ic/4+2\ic t}(2t)^\nu f_2^{(0)}\big(\sqrt{2t}\zeta(w)\big)^2 \zeta^\prime(w)H(w)e^{-t\zeta^2(w)},
\]
where
\[
H(w) = \frac{w^{5/2-\nu}}{(w+1)(w-\ic)^2}\Big(\hat{\delta}(w)^{2}\tfrac{h(w)-1}{h(w)}-\hat{\delta}(1)^{2}\tfrac{h(1)-1}{h(1)}\Big).
\]
Recall, see \eqref{alpha and beta} and \eqref{defn of s1 and s2}, that \( \hat\alpha^2 \) is equal to \( e^{-4\ic t} \) times a constant that depends only on \( \nu \). Hence, repeating the steps leading to \eqref{I-in-03}, we get for some explicit constant \( C_\nu^{(0,1)} \) that
\begin{align}
I_{in}^{(0,1)} &:=  \frac2\pi\hat\alpha^2 e^{7\pi\ic\nu/2-\pi\ic/4+2\ic t}(2t)^\nu \int_0^{\zeta_0} f_2^{(0)}\big(\sqrt{2t}\zeta\big)^2 H(w(\zeta))e^{-t\zeta^2} d\zeta \\
\label{I-in-01}
& = C_\nu^{(0,1)} e^{-2\ic t+\nu\ln t} t^{-1} + \O_\varepsilon (t^{-3/2+\varepsilon}) \qasq t\to\infty.
\end{align}
By combining \eqref{I-in-02}, \eqref{I-in-03}, and \eqref{I-in-01} we see that the first integral in \eqref{target integral ver 2} behaves like
\begin{align}
\label{I-in-0}
C_\nu^{(0,1)} e^{-2\ic t+\nu\ln t} t^{-1} + C_\nu^{(0,3)} e^{2\ic t-\nu\ln t} t^{-1} + \O_\varepsilon (t^{-3/2+\varepsilon}) \qasq t\to\infty.
\end{align}

When it comes to the second integral in \eqref{target integral ver 2} recall that  \( g^{(2)}(w,t) = e^{-2\pi\ic\nu}g^{(0)}(w,t) \) while \( \ell^{(0)}(w,t)\) and \( \ell^{(2)}(w,t) \) differ only in the choice of the functions \( f^{(i)} \). Now, we get from \eqref{defn of Z} that
\[
 f_2^{(2)}(s) = -e^{\pi\ic\nu} f_2^{(0)}(-s) \qandq f_4^{(2)}(s) = e^{\pi\ic\nu} f_4^{(0)}(-s).
 \]
 Since \( \zeta(w) \) maps \( \Gamma_2 \) into the negative reals, all our previous estimates carry over unchanged with \eqref{integal of D} becoming
 \[
 \frac1{2t}\int_0^{-\sqrt{2t}\zeta_2} sD_\nu^2(-s) ds =  \frac1{2t}\int_0^\infty sD_\nu^2(s) ds + o\big(e^{-ct}\big),
 \]
 where \( -\zeta_2 \) is the value of \( \zeta(w) \) at the point where \( \Gamma_2 \) meets \( \partial \D_\epsilon(1) \). Thus, the second integral in \eqref{target integral ver 2} behaves exactly as in \eqref{I-in-0}. Due to the minus sign in front of it, the contribution of the first two summands in \eqref{target integral ver 2} to \( I_{in} \) is of order \( \O_\varepsilon (t^{-3/2+\varepsilon}) \).

The other two integrals can be estimated similarly. We only need to observe that \( g^{(3)}(w,t) = e^{2\pi\ic\nu} g^{(1)}(w,t) \), \( \ic\zeta(w) \) maps \( \Gamma_1 \) and \( \Gamma_3 \) into the negative and positive reals while
\[
f_1^{(3)}(s) = e^{-\pi\ic\nu}f_1^{(1)}(-s) \qandq f_3^{(3)}(s) = - e^{-\pi\ic\nu} f_3^{(1)}(-s).
\]
Altogether, we get that
\begin{align}
\label{I-in}
I_{in} = \O_\varepsilon (t^{-3/2+\varepsilon}) \qasq t\to\infty \quad \text{for any} \quad \varepsilon>0.
\end{align}
Respectively, we get from \eqref{limit S1 prelim}, \eqref{I-out}, \eqref{I-bnd}, and \eqref{I-in} that
\begin{equation}
\label{limit S1}
\lim_{w \to \infty} w \big[ \mathcal{S}^{(1)}(w) \big]_{11} = d_1 e^{-2\ic t + \nu \ln t} t^{-1/2} + d_2 e^{2\ic t - \nu \ln t} t^{-1/2} + \frac{\ic \nu^2}{2t} + O_\varepsilon \big(t^{-3/2+\varepsilon}\big). 
\end{equation}

\subsection{Asymptotic Analysis of \( \mathcal S^{(2)} \)}

Next we analyze \( \mathcal{S}^{(2)}(w) = \mathcal C(\mathcal S_-^{(1)}E)(w) = \mathcal C((\mathcal C_-E)E)(w) \). We have that
\begin{align} \label{defn of wS2}
\lim_{w \to \infty} w \, \mathcal{S}^{(2)}(w) = -\frac{1}{2\pi \ic} \int_{\Gamma_R} \left( \lim_{s \to z \in \Gamma_{R-}} \frac{1}{2\pi \ic} \int_{\Gamma_R} \frac{E(w')}{w' - s} d w'\right) E(z) dz.
\end{align}
We write the integral in parentheses above as a sum of two terms, say \( J_1(s) \) and \( J_2(s) \), where \( J_1(s) \) is obtained by integrating over \( \Gamma_R^{bnd} \) and \( J_2(s) \) by integrating over \( \Gamma_R^{out}\cup\Gamma_R^{in} \). 

We start with \( J_2(s) \). Observe that \( J_{2-}(s)= \mathcal C_-(\tilde E)(s) \), where \( \tilde E(w)=E(w) \) on \( \Gamma_R^{out}\cup\Gamma_R^{in} \) and \( \tilde E(w)=0 \) on \( \Gamma_R^{bnd} \). Hence,
\[
\|J_{2-}\|_{L^2(\Gamma_R)} \leq \|\mathcal C_-\| \|\tilde E\|_{L^2(\Gamma_R)} = \O\big(t^{-3/4}\big)
\]
by \eqref{G R - I exp small outside} and \eqref{G R - I small inside}. Thus, we have that
\begin{align}
\|J_{2-}E\|_{L^1(\Gamma_R)} &\leq \|E\|_{L^\infty(\Gamma_R)}\|J_{2-}\|_{L^1(\Gamma_R)} \\
& \leq |\Gamma_R|^{1/2}\|E\|_{L^\infty(\Gamma_R)}\|J_{2-}\|_{L^2(\Gamma_R)} = \O\big(t^{-5/4}\big),
\end{align}
where we used \eqref{G R - I small} on the last step. That is, it holds that
\begin{align}
\label{limit S2 prelim}
\lim_{w \to \infty} w \big[ \mathcal{S}^{(2)}(w) \big]_{11} = - \frac1{2\pi\ic}\int_{\Gamma_R} \big[ J_{1-}(z)E(z) \big]_{11} dz + \O\big(t^{-5/4}\big)
\end{align}
as \( t\to\infty \). When it comes to \( J_1(s) \), it follows from \eqref{E symmetry} that
\begin{align} 
\label{J1}
J_1(s) = \frac{1}{2\pi \ic} \int_{\partial \D_\epsilon(1)} \frac{E(w')}{w' - s} d w' + \frac{1}{2\pi \ic} \int_{\partial \D_\epsilon(1)} \frac{\sigma_1 E(w') \sigma_1}{w' + s} d w'.
\end{align}
This formula immediately yields symmetry \( J_1(s) = \sigma_1 J_1(-s) \sigma_1 \). Denote the integral in \eqref{limit S2 prelim} by \( I \). Then,
\begin{align}
I = \frac{1}{2\pi\ic}\int_{\Gamma_R\cap\{\re(w)>0\}} \big[ \sigma_1J_{1-}(w)E(w)\sigma_1 - J_{1-}(w)E(w) \big]_{11} dw,
\end{align}
where we used \eqref{E symmetry} and the above symmetry of \( J_1(s) \). By \eqref{matching cond 1 ver 0}, it holds that
\begin{equation} 
\label{E asymptotics}
E(w) = E_1(w) + E_2(w), \quad E_1(w) = \frac{1}{\sqrt{t}} \frac{M(w)}{w-1}, \quad  E_2(w) = \O\big(t^{-1}\big),
\end{equation}
where \( M(w) \) is an analytic matrix function around \( 1 \) given by
\begin{align} 
M(w) := e^{-3\pi\ic/4}\sqrt{\frac {w}2} A(w) \begin{pmatrix}
        0 & - \tilde \alpha(w)\\
        \tilde \beta(w) & 0
    \end{pmatrix} A(w)^{-1}.
\end{align}
Write \( J_1(s) = J_{1,1}(s) + J_{1,2}(s) \), where this decomposition is based on the above decomposition \( E(w) \). Extend \( E_2(w) \) by zero to the entire contour \( \Gamma_R \). Then, similarly to the case of \( J_2(s) \), we have that
\[
\|J_{1,2-}\|_{L^2(\Gamma_R)} = \O\big(t^{-1}\big) \quad \Rightarrow  \quad \|J_{1,2-}E\|_{L^1(\Gamma_R)} = O\big(t^{-3/2}\big).
\]
Thus, we have that
\begin{align}
\label{I and I1}
I = I_1 + O\big(t^{-3/2}\big) \qasq t\to\infty,
\end{align}
where \( I_1 \) is defined exactly as \( I \), but with \( J_1(w) \) replaced by \( J_{1,1}(w) \). Now, \( J_{1,1}(s) \) can be computed explicitly. We get from the residue theorem that
\[
J_{1,1}(s) = \frac{1}{\sqrt{t}} \left( -N(s) - \frac{\sigma_1 M(1) \sigma_1}{s + 1}\right), \quad s \in \D_\epsilon(1),
\]
where \( N(s) = (M(s)-M(1))/(s - 1) \) (recall that \( \partial \D_\epsilon(1) \) and \( \partial \D_\epsilon(-1) \) are oriented clockwise). We have that
\begin{equation}
\label{defn of M}
M(w) = e^{-3\pi\ic/4}\sqrt{\frac {w}2} \begin{pmatrix} \displaystyle   \frac{w \tilde \beta(w)}{(w+\ic)^2} - \frac{w \tilde\alpha(w)}{(w-\ic)^2} & \displaystyle  - \frac{\tilde\beta(w)}{(w+\ic)^2} - \frac{w^2\tilde\alpha(w)}{(w-\ic)^2} \smallskip \\
\displaystyle  \frac{w^2\tilde\beta(w)}{(w+\ic)^2} + \frac{\tilde\alpha(w)}{(w-\ic)^2} & \displaystyle - \frac{w\tilde\beta(w)}{(w+\ic)^2} + \frac{w\tilde\alpha(w)}{(w-\ic)^2}
\end{pmatrix}.
\end{equation}
In particular, it holds that $\sigma_1 M(1) \sigma_1 = -M(1)$. Thus,
\begin{align} 
\label{J asymptotics}
J_{1,1}(s) = \frac{1}{\sqrt{t}} \left(  \frac{M(1)}{s + 1} - N(s)\right), \quad s \in \D_\epsilon(1).
\end{align}
Similarly, we get that
\begin{equation}
\label{J1 outside}
J_{1,1}(s) = \frac{1}{\sqrt{t}} \frac{2s}{s^2-1}M(1), \quad s\not\in \overline{\D_\epsilon(1) \cup \D_\epsilon(-1)}.
\end{equation}
Write \( I_1 = I_1^{out}+I_1^{bnd}+I_1^{in} \) depending on whether we integrate over \( \Gamma_R^{out} \), \( \Gamma_R^{bnd} \), or \( \Gamma_R^{in} \). We immediately get from \eqref{G R - I exp small outside} and \eqref{J1 outside} that
\begin{align}
\label{I 1 out}
I_1^{out} = \O(e^{-c_{\epsilon} t}) \qasq t\to\infty.
\end{align}
We further get from \eqref{G R - I small inside} and \eqref{J asymptotics} that
\begin{align}
\label{I 1 in}
I_1^{in} = \O\left( t^{-1/2} \|E\|_{L^1(\Gamma_R^{in})} \right) = \O\big(t^{-3/2}\big).
\end{align}
Next, it follows from \eqref{E asymptotics}, \eqref{J asymptotics}, and the relation $\sigma_1 M(1) \sigma_1 = -M(1)$ that
\begin{align}
I_1^{bnd} = \frac{1}{t}\frac{1}{2\pi\ic} \int_{\partial \D_\epsilon(1)} \big[ N(w) M(w) - \sigma_1 N(w) M(w) \sigma_1\big]_{11} \frac{dw}{w-1} + \O\big(t^{-3/2}\big).
\end{align}
We get from the residue theorem that the integral above is equal to
\begin{align}
\big[ \sigma_1 M^\prime(1) M(1) \sigma_1 - M^\prime(1) M(1) \big]_{11}  & = - \big[ \big(\sigma_1 M^\prime(1) \sigma_1 + M^\prime(1)\big) M(1) \big]_{11}  \\
& = -\big([M^\prime(1)]_{12}+[M^\prime(1)]_{21}\big)[M(1)]_{21},
\end{align}
where we used the fact that \( M(w) \) is traceless. Thus,
\begin{align}
I_1^{bnd} &= \frac{\ic}{4t} \left( \tilde \beta (1)^2 - \tilde \alpha(1)^2 \right) + \O\big(t^{-3/2}\big) \\
\label{I 1 bnd}
&= c_1 e^{4\ic t-2\nu\ln t} t^{-1} + c_2 e^{-4\ic t+2\nu\ln t} t^{-1} + \O\big(t^{-3/2}\big) \qasq t\to\infty
\end{align} 
for some explicit constants \( c_1, c_2 \), where one needs to recall \eqref{defn of s1 and s2}, \eqref{alpha and beta}, and \eqref{alpha tilde and beta tilde}.   Gathering together \eqref{limit S2 prelim},  \eqref{I and I1}, \eqref{I 1 out}, \eqref{I 1 in}, and \eqref{I 1 bnd}, we get that 
\begin{equation}
\label{limit S2}
\lim_{w \to \infty} w \big[ \mathcal{S}^{(2)}(w) \big]_{11} = c_1 e^{-4\ic t+2\nu\ln t} t^{-1} + c_2 e^{4\ic t-2\nu\ln t} t^{-1} + \O\big(t^{-5/4}\big) \qasq t\to\infty.
\end{equation}


\subsection{Proof of Theorem~\ref{thm:main}}

From the very definition of \( \mathcal S^{(k)}(w) \) we get that
\begin{align}
\left|\lim_{w\to\infty} w\mathcal S^{(k)}(w) \right| \leq \frac1{2\pi} \|\mathcal S_-^{(k-1)} E\|_{L^1(\Gamma_R)} \leq \frac{|\Gamma_R|^{1/2} }{2\pi} \| E \|_{L^\infty(\Gamma_R)} \|\mathcal S_-^{(k-1)}\|_{L^2(\Gamma_R)}.
\end{align}
Iterating the last estimate gives
\begin{align}
\left|\lim_{w\to\infty} w\mathcal S^{(k)}(w) \right| \leq \frac{|\Gamma_R|^{1/2}}{2\pi} \|I\|_{L^2(\Gamma_R)}  \|\mathcal C_-\|^{k-1} \| E \|_{L^\infty(\Gamma_R)}^k.
\end{align}
Hence, we have that
\begin{align}
\sum_{k=3}^\infty\left|\lim_{w\to\infty} w\mathcal S^{(k)}(w) \right| \leq \frac{|\Gamma_R|^{1/2}}{2\pi} \frac{\|I\|_{L^2(\Gamma_R)}}{\|\mathcal C_-\|} \frac{\|\mathcal C_-\|^3 \| E \|_{L^\infty(\Gamma_R)}^3 }{1-\|\mathcal C_-\| \| E \|_{L^\infty(\Gamma_R)}} = \O\big( t^{-3/2} \big)
\end{align}
as \( t\to\infty \), where the last conclusion follows from \eqref{G R - I small}.  Thus, we get from \eqref{R1 11 component eq 1}, \eqref{limit S1}, and \eqref{limit S2} that
\begin{multline}
[R_1(t)]_{11} =  d_1 e^{-2\ic t + \nu \ln t} t^{-1/2} + d_2 e^{2\ic t - \nu \ln t} t^{-1/2} + c_1 e^{-4\ic t+2\nu\ln t} t^{-1} \\
+ c_2 e^{4\ic t-2\nu\ln t} t^{-1} + \frac{\ic \nu^2}{2t} + \O\big( t^{-5/4} \big) \qasq t\to\infty.
\end{multline}
It follows from \eqref{P global} and \eqref{defn of R} that
\[
[Y_1(t)]_{11} = [R_1(t)]_{11} + \ic.
\]
Therefore, we get from \eqref{defn of delta 1} and \eqref{sigma and Y1} that
\[
\sigma^\prime(t) = t + \frac{1}{\pi} \int_{-1}^1 \ln |\tanh \beta \zeta| d\zeta  + \ic [R_1(t)]_{11}. 
\]
Observe that for a given complex number \( u \) with \( \re(u)>0 \) and a real number \( x \), it holds that
\begin{align}
\ic x\int_1^t s^{-u} e^{\ic xs}ds & = t^{-u}e^{\ic xt} - e^{\ic x} + u\int_1^t s^{-u-1}e^{\ic xs}ds \\
& = u\int_1^\infty s^{-u-1}e^{\ic xs}ds - e^{\ic x} + \O\big( t^{-\re(u)}\big) \qasq t\to\infty
\end{align}
because the last integral above is absolutely convergent. Hence, we get that
\[
\sigma(t) = \frac12 t^2 + \frac{t}{\pi} \int_{-1}^1 \ln |\tanh \beta \zeta| d\zeta - \frac{\nu^2}2 \ln t + C + \O\big(t^{-1/4}\big)
\]
as \( t\to\infty \) for some constant \( C \). The claim of Theorem~\ref{thm:main} now follows from \eqref{defn of sigma}, the definition of \( \nu \) in \eqref{defn of nu and h hat}, and the definition of \( h(w) \) in \eqref{defn of h}.

\bibliographystyle{alpha}
\bibliography{reference.bib}

\end{document}